\documentclass[aps,prx,twocolumn,preprintnumbers,superscriptaddress,longbibliography]{revtex4-2}
\usepackage{amsmath}
\usepackage{amssymb}
\usepackage{enumitem}
\usepackage{color}
\usepackage{xcolor}
\usepackage{graphicx}
\usepackage{dcolumn}
\usepackage{bm}
\usepackage[colorlinks=true,linkcolor=blue,citecolor=blue,urlcolor=blue, hypertexnames=false]{hyperref}
\usepackage{feynmp-auto}    
\usepackage{ulem}           
\newcommand{\drop}[1]{}

\definecolor{BluBondi}{rgb}{0.00,0.58,0.71}
\definecolor{tangerine}{rgb}{0.944,0.522,0}
\definecolor{brown}{rgb}{0.633,0.156,0.156}
\definecolor{byzantine}{rgb}{0.74, 0.2, 0.64}
\definecolor{teal}{rgb}{0,0.502,0.502}
\newcommand{\editor}[2]{%
  \expandafter\newcommand\csname #1note\endcsname[1]{%
    \textcolor{#2}{(\textbf{#1:} \textit{##1})}}%
  \expandafter\newcommand\csname #1\endcsname[1]{%
    \textcolor{#2}{##1}}%
  \expandafter\newcommand\csname #1cancel\endcsname[1]{%
    \textcolor{#2}{\sout{##1}}}%
  \expandafter\newcommand\csname #1change\endcsname[2]{%
    \textcolor{#2}{\sout{##1} ##2}}%
  \newenvironment{#1text}{\color{#2}}{\color{black}}
}
\editor{AF}{BluBondi}
\editor{TC}{tangerine}
\editor{MQ}{red}
\editor{MG}{teal}
\newcommand{\Trw}{\text{Tr}_\omega}
\newcommand{\TrLn}{\text{Tr}_\omega\text{Ln}}
\newcommand{\ket}[1]{|#1\rangle}

\newcommand{\braket}[2]{\langle #1 | #2 \rangle}
\makeatletter
\def\maketitle{
\@author@finish
\title@column\titleblock@produce
\suppressfloats[t]}
\makeatother
\begin{document}
%
%====== TITLE===================================
\title{Broken symmetry solutions in one-dimensional lattice models\\ via many-body perturbation theory}
%
%====== AUTHORS ================================
\author{Matteo Quinzi}
\email[corresponding author: ]{matteo.quinzi@epfl.ch}
\affiliation{Dipartimento di Scienze Fisiche, Informatiche e Matematiche, Universit\`a
degli Studi di Modena e Reggio Emilia, Via G.~Campi 213/a, 41125 Modena, Italy}
\affiliation{Theory and Simulations of Materials (THEOS) and National Centre for Computational Design and Discovery of Novel Materials (MARVEL), \'Ecole Polytechnique F\'ed\'erale de Lausanne, 1015 Lausanne, Switzerland}
\author{Tommaso Chiarotti}
\affiliation{Theory and Simulations of Materials (THEOS) and National Centre for Computational Design and Discovery of Novel Materials (MARVEL), \'Ecole Polytechnique F\'ed\'erale de Lausanne, 1015 Lausanne, Switzerland}
\author{Marco Gibertini}
\affiliation{Dipartimento di Scienze Fisiche, Informatiche e Matematiche, Universit\`a
degli Studi di Modena e Reggio Emilia, Via G.~Campi 213/a, 41125 Modena, Italy}
\affiliation{Centro S3, CNR--Istituto Nanoscienze, 41125 Modena, Italy}
\author{Andrea Ferretti}
\affiliation{Centro S3, CNR--Istituto Nanoscienze, 41125 Modena, Italy}
\affiliation{Dipartimento di Scienze Fisiche, Informatiche e Matematiche, Universit\`a
degli Studi di Modena e Reggio Emilia, Via G.~Campi 213/a, 41125 Modena, Italy}
\date{\today}
%
%========= ABSTRACT =============================
\begin{abstract}
In this work we study self-consistent solutions in one-dimensional lattice models obtained via many-body perturbation theory.
The Dyson equation is solved in a fully self-consistent manner via the algorithmic-inversion method based on the sum-over-poles representation (AIM-SOP) of dynamical operators.
In particular, we focus on the GW approximation, analyzing the spectral properties and the emergence of possible magnetic- or charge-density-wave
broken symmetry solutions.
We start by validating our self-consistent AIM-SOP implementation by taking as test case the one-dimensional Hubbard model.
We then move to the study of antiferromagnetic  and charge density wave solutions  in one-dimensional lattice models, taking into account a long-range Coulomb interaction between the electrons.
We show that moving from local to non-local electronic interactions leads to a competition between antiferromagnetic and charge-density-wave broken symmetry solutions.
Complementary, by solving the Sham-Schl\"uter equation, we can compute the non-interacting Green's function reproducing the same charge density of the interacting system.
In turn, this allows for the
evaluation of the derivative discontinuity of the Kohn-Sham (KS) potential, showing that its contribution to the fundamental gap can become dominating in some of the studied cases.
\end{abstract}
%
%================================================
%         PAPER MAIN BODY
%================================================
%
\maketitle
%
%======== INTRODUCTION ==========================
%
\section{Introduction}
\label{sec:intro}
One-dimensional systems constitute an optimal playground to study the effect of electronic correlations at a reduced computational cost with respect to systems of higher dimensionality.
A recent study of one-dimensional hydrogen chains~\cite{Motta_PRX_2020} has shown how the presence of long-range Coulomb repulsion between the electrons can give rise to a rich phase diagram, including antiferromagnetic (AFM) domains and metal-to-insulator (MIT) transitions.
Furthermore one-dimensional quasi-metals constitute a prototypical example of Peierls transitions~\cite{Peierls_Annalen_1929}, where a periodic lattice distortion destabilizes the metallic phase leading to an insulating state with a lower total energy.

However it has been recently pointed out that the same systems can host a purely electronic charge density wave (CDW) that does not involve a structural symmetry breaking~\cite{Barborini_PRB_2022}. 
This electronic CDW, which destabilizes the one-dimensional quasi-metal by opening a small energy gap could be related to an excitonic mechanism~\cite{SherringtonKohn_RevModPhys_1968, Varsano_Nature_2017}.
However studies performed on one-dimensional Hubbard chains~\cite{Hubbard_ProcLondon_1963} have shown that an approximate treatment of the electron-electron repulsion can lead to unphysical solutions with a broken symmetry ground state \cite{Joost2021ContrPlasmaPhys, Honet2022AIP}: this is the case of antiferromagnetic (AFM) solutions found via many-body perturbation theory~\cite{fett-wale71book,Martin-Reining-Ceperley2016book,Stefanucci-vanLeeuwen2013book}, while an analytical treatment of one-dimensional Hubbard clusters predicts a non-magnetic (NM) ground state~\cite{Lieb_Hubb_1968PRL,Lieb_Hubb_2003_Phys.A} (also referred to as a paramagnetic state in the literature~\cite{Honet2022AIP}).

In the framework of MBPT, the information about the charged excitations of the interacting system is encoded in the one-particle Green's function (GF), with the effect of the electronic interaction described by the non-local and dynamical self-energy operator $\Sigma(\omega)$. 
For conserving approximations~\cite{Baym1961PhysRev,Baym1962PhysRev}, the MBPT problem can also be recast in a functional formulation via the Luttinger-Ward~\cite{Luttinger1960PhysRev} and Klein~\cite{klein_PhysRev_1961} functionals, which ultimately require the Dyson equation to be solved in a self-consistent way.
Despite the potential of MBPT, density functional theory (DFT)~\cite{HohenbergKohn_PhysRev_1964} in its Kohn-Sham formulation (KS-DFT)~\cite{KohnSham_PhysRev_1965} is still one of the mostly used tools in electronic structure calculations~\cite{Marzari_NatureMaterials_2021}.
The bridge between the MBPT framework and the KS-DFT formalism is given by the Sham-Schl\"uter equation (SSE)~\cite{Sham_PRB_1985,ShamSchluter_PRL_1983} that allows one to compute the local Kohn-Sham exchange-correlation potential once the self-consistent GF and its corresponding self-energy are known.
Although the SSE can be solved self-consistently, the solution is typically found in its linearized form~\cite{Casida_PRA_1995}.

Part of the difficulty in solving the self-consistent Dyson equations stems from the handling of dynamical potentials. In this respect, Chiarotti et al.~\cite{Chiarotti2022PRR,Chiarotti2023PhD,Chiarotti2024PRR} have recently exploited sum-over-poles (SOP),
i.e.\ meromorphic (and rational) functions of the frequency having only simple (first order) poles, to represent dynamical propagators and self-energies. Similar constructions have  also been employed, e.g., in Refs.~\cite{Engel1991PRB,Friesen2010PhysRevB,Sabatino2021FrontChem}. 
Moreover, in Refs.~\cite{Chiarotti2022PRR,Chiarotti2023PhD,Chiarotti2024PRR},
the SOP representation is combined with the algorithmic-inversion method (AIM-SOP) providing an in-principle exact way of solving the Dyson equation at all frequencies by means of the diagonalization of a linear eigenvalue problem in a larger space~\cite{Guttel2017ActaNumerica}.
Similar approaches have been used in the context of DMFT, GW, Bethe-Salpeter equation~\cite{Savrasov2006PhysRevLett,Budich2012PhysRevB,Wang2012PhysRevB,Bintrim2021JChemPhys,Bintrim2022JChemPhys, MoninoLoos_JChemPhys_2023}.
For instance, SOP-representations have been recently exploited in the multi-pole approximation (MPA) for full-frequency GW calculations~\cite{Leon2021PhysRevB,Leon2023PhysRevB,Leon2025arXiv} to represent the RPA response functions, self-energies, and Green's functions. 
In addition, 
the AIM-SOP framework can also be exploited~\cite{Ferretti2024PRB} to provide a fully analytical way of evaluating the Klein functional~\cite{klein_PhysRev_1961}.

In this work we study self-consistent solutions of the Dyson equation as resulting from different self-energy approximations in a MBPT context, applied to one-dimensional lattice models, also including long-range interactions beyond the local Hubbard repulsion.
Additionally, we also solve the -- non-linearized -- SSE and evaluate the derivative discontinuity of the Kohn-Sham potential. 
Technically, we make use of the AIM-SOP framework~\cite{Chiarotti2022PRR,Chiarotti2024PRR,Chiarotti2023PhD} in order to implement a fully self-consistent solution of the Dyson equation, and also exploit a recently developed analytically expression to evaluate the $\Phi[G]$ terms of the Klein functional~\cite{Ferretti2024PRB}.
Our results show that the AIM-SOP framework is capable of solving self-consistently the Dyson equation in a stable and accurate way for the systems considered, obtaining an excellent agreement with Refs.~\cite{Friesen2010PhysRevB,Sabatino2021FrontChem,Joost2021ContrPlasmaPhys}.

Physically, when a long-range electronic repulsion is introduced in the model, both AFM and CDW solutions can be found depending on the strength of the non-local interaction term, consistently with the results of Refs.~\cite{Motta_PRX_2020,Barborini_PRB_2022}.
We show that the interplay between the local and  non-local electronic interactions leads to a competition between AFM and CDW phases in the exact ground state, and thus to an explicit symmetry breaking in the case of MBPT solutions. 
This effect is particularly relevant in the case of a non-interacting metallic ground state.
Remarkably, the evaluation of the total energy within the GW approximation improves considerably in the case of symmetry unrestricted calculations.
Furthermore, by solving the SSE problem, we also find that the KS potential displays a significant derivative discontinuity when CDW solutions are found, in agreement with Ref.~\cite{SchonhammerGunnarsson_JournalOfPhysicsC_1987}, stressing the role of exchange-correlation corrections to the fundamental gap of the KS system.

This manuscript is organized as follows: in Sec.~\ref{sec:framework} we give an overview of the MBPT formalism. In Sec.~\ref{sec:sop_aim} we explain how the Dyson equation is solved in a fully self-consistent way via the AIM-SOP procedure. In Sec.~\ref{sec:validation} we test the accuracy and performance of the AIM-SOP procedure using one-dimensional Hubbard chains as test case and we compare our results against those of other authors. In Sec.~\ref{sec:long-range} we study the presence of broken symmetry (AFM and CDW) solutions in periodic one-dimensional lattices in presence of a non-local electron-electron repulsion. Additional and complementary details are reported in the Appendixes and in the Supplemental Material~\cite{suppinfo}
(see also references~\cite{Joost2021ContrPlasmaPhys, Romaniello_hubGW_2009J.Chem.Phys, Ferretti2024PRB, Stan2015NewJPhys, GalitskiiMigdal_SovPhys_1958, farid2021luttingerwardfunctionalconvergenceskeleton} therein).
%
%========= FRAMEWORK ============================
%
\section{Theoretical Framework}
\label{sec:framework}
%
% -- MBPT --
\subsection{MBPT Theory}
\label{subsec:mbpt_theory}
The one-body Green's function~\cite{fett-wale71book,Stefanucci-vanLeeuwen2013book,Martin-Reining-Ceperley2016book} is a dynamical operator that allows one to evaluate expectation values of one-body observables, charged excitations (one particle spectrum), and total energies, e.g. through the Galitskii-Migdal (GM) formula~\cite{GalitskiiMigdal_SovPhys_1958} or the Luttinger-Ward and Klein functionals~\cite{Luttinger1960PhysRev,klein_PhysRev_1961}.
In this work we focus on zero-temperature systems at equilibrium.
Starting from a non-relativistic independent-particle Hamiltonian $h_0$, with the non-interacting Green's function $G_0$ written as
\begin{equation}
    G_0(\omega) = \left[\omega I - h_0\right]^{-1},
\end{equation}
the effect of the electron-electron interactions is taken into account through the introduction of the self-energy $\Sigma(\omega)$, which is in general a non-local and dynamical operator. 
The GF of the interacting system is then obtained from the Dyson equation
\begin{equation}
\label{eq:dyson}
    G(\omega) = \left[\omega I - h_0 - \Sigma(\omega)\right]^{-1}.
\end{equation}

Approximations to the self-energy (SE) operator can be obtained, e.g., within MBPT, where the SE can be diagrammatically defined by considering selected classes of diagrams.
In this regard, Baym and Kadanoff~\cite{Baym1962PhysRev,Baym1961PhysRev} have shown that conserving approximations for $\Sigma=\Sigma[G]$ are obtained when one can introduce a functional $\Phi[G]$ such that
\begin{equation}
\label{eq:sgm_from_phi}
    \Sigma[G] = {2\pi i}\frac{\delta \Phi[G]}{\delta G}.
\end{equation}
The $\Phi[G]$ term can then be used within the Luttinger-Ward  \cite{Luttinger1960PhysRev} or Klein~\cite{klein_PhysRev_1961} functionals to express the total energy of the system.
We will focus on the latter, which is given in terms of $G$ by:  
\begin{align}
\label{eq:klein_functional}
    E_{K}[G] = &\TrLn\left\{G_0^{-1}G\right\} 
                    + \Trw\left\{h_0 G_0\right\} \\
    \nonumber 
                    &+ \Trw\left\{I - G_0^{-1}G\right\}
                    + \Phi[G],
\end{align}
where we use the notation $\Trw [\dots] =\int \frac{d\omega}{2\pi i}e^{i\omega 0^+}\text{Tr}[\dots]$.
While these functionals assume in general different values~\cite{Dahlen_PhysRevA.73.012511_2006, Lani2024PhysRevRes} for a generic $G$, they both return the ground state total energy at their stationary points.
Notably, the $G$ that makes the functional stationary is obtained by solving a self-consistent Dyson equation, where the SE depends on $G$ via Eq.~\eqref{eq:sgm_from_phi}.
It is also helpful to stress that the $\Phi$ functional can be built through a non-diagrammatic and non-perturbative procedure~\cite{Potthoff2006CondensMatPhys}, thereby being well suited to study broken symmetry solutions of systems in the strong correlation regime. 
The self-energy approximations considered in this work are the Hartree-Fock (HF), second Born (2B), and GW approximations.
Their SOP representation is reported in Appendix~\ref{app:sop_self_ene}.
%
%-- SHAM SCHLUTER EQUATION --
%
\subsection{Bridging MBPT and KS-DFT frameworks}
\label{subsec:sse_theory}
As stated in Sec.~\ref{sec:intro}, it is possible to bridge MBPT and KS-DFT via the SSE.
One defines the KS GF through the use of a local and static potential $v^{\text{KS}}(\mathbf{r})$ via
\begin{equation}
\label{eq:G_ks}
    G^\text{KS}(\omega) = \left[\omega I - h_0 - v^\text{KS}\right]^{-1}.
\end{equation}
The local potential $v^\text{KS}(\mathbf{r})$ is determined in such a way to reproduce the same charge density of the interacting system.
This is obtained by solving the following Sham-Schl\"uter equation (SSE) \cite{ShamSchluter_PRL_1983,Sham_PRB_1985}:
\begin{align}
    \label{eq:sse}
    \nonumber 
    \int d\mathbf{r}_1 \int \frac{d\omega}{2\pi i}e^{i\omega 0^+}& G^\text{KS}(\mathbf{r},\mathbf{r}_1;\omega) \, v^\text{KS}(\mathbf{r}_1) \, G(\mathbf{r}_1,\mathbf{r};\omega) = \\
    \nonumber 
    \int d\mathbf{r}_1 d\mathbf{r}_2 &\int \frac{d\omega}{2\pi i}e^{i\omega 0^+}\Big[ G^\text{KS}(\mathbf{r},\mathbf{r}_1;\omega) \times \\
    &\Sigma[G](\mathbf{r}_1,\mathbf{r}_2;\omega)G(\mathbf{r}_2,\mathbf{r};\omega)\Big],
\end{align}
where the dependence of the SE on the self-consistent GF is highlighted.
Since the potential $v^\text{KS}$  
is defined up to a constant, corresponding to a rigid energy shift leaving the density unchanged, we require $G^\text{KS}$ and $G$ to share the same Fermi level.

Furthermore the evaluation of $G^\text{KS}$ through Eq.~\eqref{eq:G_ks} allows one to determine the value of the discontinuity $\Delta_{xc}$ in the KS potential for a given SE approximation at the MBPT level~\cite{ShamSchluter_PRL_1983,GodbyShamScluter_PRL_1986, Farid_PhysRevB.38.7530_1988}.
This space-independent quantity relates the energy gap $E^\text{KS}_g$ of the KS system to the physical fundamental gap, i.e.\ the quasiparticle energy gap $E^\text{QP}$ evaluated at the MBPT level
\begin{equation}
    \label{eq:delta_xc}
    E^\text{QP}_g - E^\text{KS}_g = \Delta_{xc} = v^{\text{KS}}_{+} - v_{-}^\text{KS},
\end{equation}
with $v_{\pm}^\text{KS}$ being the KS potentials associated to an infinitesimal increase or decrease in the number of particles.

In passing, we notice that a linearized version of the SSE, where the self-consistent $G$ is replaced by $G^\text{KS}$ everywhere (also in the self-energy dependence) has been originally proposed by Casida~\cite{Casida_PRA_1995} and it is used in the context of optimized effective potential (OEP) approaches~\cite{Kummel2008RevModPhys,Martin-Reining-Ceperley2016book,Ferretti_PRB_2014,Hellgren2018PhysRevB,Vacondio_JChemTheoryComput_2022}. 
A dynamical generalization of the SSE has also been proposed by  Gatti et al.~\cite{Gatti_PRL_2007} and it has been shown to be flexible enough to reproduce the spectral density of interacting systems~\cite{Gatti_PRL_2007,Ferretti_PRB_2014,Vanzini2018TheEPJB,Vanzini2020FaradayDiscuss}.
The SOP expression for the linear system defined by Eq.~\eqref{eq:sse} is reported in Appendix~\ref{sec:sse_sop}.
%
% ========== METHODOLOGY =========================
%
\section{The AIM-SOP approach}
\label{sec:sop_aim}
%
% -- AIM-SOP --
\subsection{Sum-over-pole and algorithmic-inversion method}
In the context of Green's function methods, the SOP representation of propagators or dynamical operators is a meromorphic representation, typically involving only simple poles. 
This has been recently exploited in Ref.~\cite{Chiarotti2022PRR} to study the homogeneous electron gas, and also used 
in a first-principles context in combination with a localized GW formulation~\cite{Chiarotti2023PhD,Chiarotti2024PRR}.  Analogous meromorphic constructions have also been employed elsewhere when dealing with Green's function methods~\cite{Verdozzi1987NuovoCimentoD, Engel1991PRB,Friesen2010PhysRevB,Sabatino2021FrontChem}. 
In particular, the work of Engel and Farid~\cite{Engel1991PRB} shows how the SOP representation can also be seen as a truncated Lehmann representation.
In Refs.~\cite{Chiarotti2022PRR,Chiarotti2023PhD,Chiarotti2024PRR} the SOP representation is combined with an algorithmic-inversion method (AIM-SOP) to solve the Dyson equation exactly and at all frequencies. 
In the AIM-SOP approach the non-linear eigenvalue problem (NLEP) described by the Dyson equation is recast into a linear eigenvalue problem in a larger space. 
The same AIM-SOP approach allows one to evaluate analytically the Klein functional and the RPA correlation energy~\cite{Ferretti2024PRB}.
In this section, based on the work of Refs.~\cite{Chiarotti2022PRR,Chiarotti2024PRR,Ferretti2024PRB}, we give a self-contained description of the AIM-SOP framework. 

Adopting a lattice representation, which is the focus of this work, with $i$ and $j$ denoting site indices, the interacting GF and the self-energy are expressed as 
\begin{align}
    %\label{eq:sop_g0}
    %G_{0,ij}(\omega) &= \sum_n \frac{A^{0,n}_{ij}}{\omega - z_n},\\
    %
    \label{eq:sop_g}
    G_{ij}(\omega) &= \sum_s \frac{A^s_{ij}}{\omega - z_s},\\
    \label{eq:sop_sgm}
    \Sigma_{ij}(\omega) &= \Sigma^0_{ij} + \sum_m \frac{\Gamma^m_{ij}}{\omega - \Omega_m},
\end{align}
with simple poles $z_s$ and $\Omega_m$, and residues defined by the matrix elements $A^s_{ij}$ and $\Gamma^m_{ij}$, respectively.
Regarding the GF, the SOP representation then coincides with the Lehmann representation~\cite{Lehmann_NuovoCimento_1954} of a dynamical propagator of a finite system with discrete one-particle excitations, for which the singularities of the GF are isolated and do not merge into a branch cut on the real axis.  
In passing we stress that, at variance with Refs.~\cite{Chiarotti2022PRR,Chiarotti2023PhD,Chiarotti2024PRR}, we only consider propagators having real poles, while the AIM-SOP framework allows one to treat the general case of complex valued poles.
With real poles, the time-ordering of propagators can be defined once a Fermi level $\mu$ is specified~\cite{fett-wale71book,Martin-Reining-Ceperley2016book,Stefanucci-vanLeeuwen2013book,Farid_Luttrev_PhilMagB1999}, by setting, e.g. for the Green's function, $z_s = \epsilon_s + i\,\text{sgn}(\mu-\epsilon_s)\eta$, with $\epsilon_s \in \mathbb{R}$
and $\eta \rightarrow 0^+$. 

The advantage of working with a meromorphic representation of dynamical operators is that several quantities of interest can be evaluated analytically.
For instance the matrix elements of the one-body reduced density matrix $\gamma_{ij}$ can be determined as
\begin{equation}
    \label{eq:density_mat_sop}
    \gamma_{ij} = \int \frac{d\omega}{2\pi i}e^{i\omega 0^+} G_{ij}(\omega) = \sum^{occ}_s A^s_{ij},
\end{equation}
where the sum is restricted to the ``occupied" poles of the dynamical operators -- i.e. the poles at energies below the Fermi level. 
Similar expressions are easily obtained for the spectral function $A_{ij}(\omega)$~\cite{fett-wale71book,Martin-Reining-Ceperley2016book,Stefanucci-vanLeeuwen2013book} and for the Galitskii-Migdal formula~\cite{GalitskiiMigdal_SovPhys_1958}.
In addition, a dynamical convolution of two SOP operators is still a SOP operator \cite{Chiarotti2022PRR} and its structure in terms of poles and residues can be determined analytically.
As stated before, and following Refs.~\cite{Farid_Luttrev_PhilMagB1999,Onida_reviewDFT-GF_RevModPhys2002,Chiarotti2022PRR,Chiarotti2024PRR}, the problem of solving the Dyson equation, Eq.~\eqref{eq:dyson}, is equivalent to the following non-linear eigenvalue problem (NLEP):
\begin{align}
\label{eq:nlep}
    \left[z_s I - h_0 - \Sigma(z_s)\right]\ket{f_s} = 0,
\end{align}
where the eigenvalues $z_s$ give the poles of the GF, while the eigenvectors $\ket{f_s}$ correspond to the Dyson orbitals and the matrix elements of their projectors give the residues of the GF at the corresponding pole, $A^s_{ij}~=~\braket{i}{f_s}\braket{f_s}{j}$ (here written in the case of a Hermitian NLEP).
For a generic operator $\Sigma(z)$, the eigenvalues defined by the NLEP of Eq.~\eqref{eq:nlep} will be in general complex valued~\cite{Guttel2017ActaNumerica}, and left and right eigenvectors need to be introduced.

The NLEP in the one-particle space can however be recast into a linear eigenvalue problem in a higher dimensional space. 
This is done by considering first a factorization of the residues of the self-energy as $\Gamma^m = V_m V^\dagger_m$, 
\begin{equation}
\label{eq:sgm_unfold}
    G^{-1}(\omega) = \omega I - h_0 -\Sigma_0 -\sum_m V_m \frac{1}{\omega - \Omega_m} V^\dagger_m,
\end{equation}
where, for a physical self-energy, the above factorization is always possible due to the positive semi-definiteness (PSD) of the residues $\Gamma^m$~\cite{Stefanucci-vanLeeuwen2013book}.
The GF of Eq.~\eqref{eq:sgm_unfold} can then be interpreted as the GF of a system coupled to a non-interacting bath: the poles of the self-energy give the degrees of freedom of the non-interacting bath, while the operators $V_m$ and $V^\dagger_m$ provide the coupling between the bath and the one-particle system. 
The diagonalization of the overall system-plus-bath Hamiltonian returns the poles of the GF as eigenvalues and the Dyson orbitals as the projection of the eigenvectors on the system subspace~\cite{Chiarotti2024PRR}.

Finally the AIM-SOP framework allows one to evaluate the Klein functional of Eq.~\eqref{eq:klein_functional} in a fully analytical manner~\cite{Ferretti2024PRB}. 
In particular, all the expressions involving the $\TrLn$ term can be computed as
\begin{align}
    \label{eq:TrLn}
    \nonumber 
    \TrLn&\left\{G_0^{-1} G\right\} = \\ &\sum_s^{occ} r_s z_s - \left[\sum_m^{occ} r_m \Omega_m + \sum_n^{occ} r_n z^0_n\right],
\end{align}
where the sum is restricted over the ``occupied" poles and each pole is weighted with its corresponding rank $r_i$ (i.e., its degeneracy).
In general, the expression in Eq.~\eqref{eq:TrLn} can be used to evaluate the $\TrLn$ term involving any two dynamical operators linked by a Dyson equation, as in Eq.~\eqref{eq:dyson}, with a kernel $\Sigma$ having at most simple poles $\Omega_m$.
This is used in practice to evaluate the RPA correlation energy of the GW approximation as discussed in Ref.~\cite{Ferretti2024PRB}.
It is relevant to stress that the poles appearing in Eq.~\eqref{eq:TrLn} are weighted with their ranks, and not with their residues: this means that poles with limited weight, such as those from satellites, count as much as quasiparticle poles in the evaluation of the total energy. 
In this respect, numerical accuracy and consistency in the solution of the Dyson equation are critical when evaluating Eq.~\eqref{eq:TrLn}.
Moreover, having a stable value from the evaluation of Eq.~\eqref{eq:TrLn} can be taken as a sign of having a good resolution in the satellite poles.
%
% -- SELF-CONSISTENCY --
%
\subsection{Self-consistency}
\label{sec:self_consistency}
Since the Dyson equation is solved iteratively, the accuracy of the algorithm is crucial to obtain faithful solutions of the self-consistent cycle.
Considering a non-interacting $G_0$ starting point, at each iteration the self-energy is evaluated in terms of its SOP representation and the Dyson equation is solved via the algorithmic-inversion method described in Sec.~\ref{sec:sop_aim}. 
The self-consistency loop is stopped whenever each matrix element $\gamma_{ij}$ of the one-body density matrix changes less than a given threshold $\epsilon$ between two successive iterations:
\begin{equation}
    \left|\gamma^{out}_{ij} - \gamma^{in}_{ij}\right| < \epsilon.
\end{equation}
Typical thresholds are $10^{-8} < \epsilon < 10^{-6}$.

In order to keep the number of poles under control we adopt a decimation procedure consisting in merging and pruning the poles of the propagators or self-energies, which is commonly employed when working with the meromorphic representation of dynamical operators~\cite{Friesen2010PhysRevB,Sabatino2021FrontChem}.
If the distance between two poles is smaller than a given threshold $\delta_M$, the poles are merged into a new pole, with the residue given by the sum of the residues of the former poles, and located at a weighted average of the previous poles. The  weights $w_s$ are related to the trace of the residues by
\begin{equation}
\label{eq:res_weights}
    w_s = \left| \text{Re} \, \text{Tr}\!\left\{A^s\right\} \right|,
\end{equation}
where the $A^s$ are the residues of the dynamical operator as in Eq.~\eqref{eq:sop_g}.
If all the matrix elements of the residue are instead smaller than the pruning threshold $\delta_P$, the pole is then eliminated and its residue is redistributed among the two nearest non-pruned poles, with weights $w_s$ given again by Eq.~\eqref{eq:res_weights}.
Since the pruning and merging operations may alter the value of the ranks of the residues, the $\TrLn$ term appearing in Eq.~\eqref{eq:TrLn} is always evaluated before these operations are performed.

In order to achieve self-consistency, it is generally necessary to mix dynamical operators calculated at different iterations of the self-consistency loop.
A commonly adopted strategy is that of linearly mixing the GFs between two successive iterations~\cite{Friesen2010PhysRevB,Joost2021ContrPlasmaPhys,Honet2022AIP}. 
Alternatively, we also adopt a linear mixing between the self-energies coming from two successive iterations~\cite{Koval2014PRB}:
\begin{align}
    \Sigma^{mix}(\omega) &= \alpha \Sigma^{out}(\omega) + (1-\alpha)\Sigma^{in}(\omega)\\
    G^{mix}(\omega) &= \left[\omega I - h_0 - \Sigma^{mix}(\omega)\right]^{-1},
\end{align}
where the mixing parameter is $0<\alpha<1$ and the mixed GF is then evaluated by solving a Dyson equation with the mixed self-energy.
According to our simulations, the $G$-mixing scheme tends to converge faster than $\Sigma$-mixing, at least for the systems studied in this work.

The Sham-Schl\"uter equation, Eq.~\eqref{eq:sse}, is solved in practice after the solution of the self-consistent Dyson equation.
The SSE on a discrete lattice defines a linear system of equations having $v^\text{KS}(\mathbf{r})$ as unknown.
This is solved self-consistently by starting from a guess of the KS potential (e.g., the local entries of the static part of the electron-electron self-energy) and solving the system by iterations until self-consistency in the density matrix is obtained.
At each iteration of the SSE loop the highest ``occupied" pole of $G^\text{KS}$ is aligned to that of the corresponding $G$ to ensure that they share the same Fermi level. 
%
%
%============ VALIDATION ========================
%
\section{Validation}
\label{sec:validation}
We start by comparing the results of our work against the available literature. 
In particular, the Hubbard model constitutes an optimal playground to study the effects of electronic correlation, where self-consistent MBPT calculations have already been performed.
The Hubbard model on a lattice~\cite{Hubbard_ProcLondon_1963, Gutzwiller_PRL_1963, Kanamori__ProgTheo_1963} includes only a local electron-electron repulsion beyond the independent-particle hopping and its Hamiltonian is given by 
\begin{equation}
    \label{eq:hubbard_hamiltonian}
    H = - t \sum_\sigma \sum_{\langle i,j \rangle} 
    \left(c^\dagger_{i\sigma } c_{j\sigma} + h.c.\right)
    + U\sum_i n_{i\uparrow }n_{i\downarrow},
\end{equation}
where $\sigma$ is the spin index with $\sigma \in [\uparrow,\downarrow]$, $t$ is the hopping parameter describing the hopping amplitude between two nearest neighboring sites (as indicated by the angular brackets in the expression above) and $U$ is the on-site electron-electron repulsion term. 
The operators $c^\dagger_{i\sigma}$ and $c_{i\sigma}$, respectively, create and destroy an electron with spin $\sigma$ on the lattice site $i$. 

In this work we focus on the one-dimensional Hubbard model with only one orbital per site  and at half-filling.
We compare our results to those in Refs.~\cite{Friesen2010PhysRevB,Sabatino2021FrontChem,Joost2021ContrPlasmaPhys}.
To this aim we stress that, in the case of small Hubbard chains, an exact treatment is also numerically feasible through exact diagonalization, which is reported in the Supplemental Material~\cite{suppinfo}.
An exact analytical treatment of the system also shows that no metal-insulator transition (MIT) is found in the case of chains with an even number of sites and at half-filling~\cite{Lieb_Hubb_1968PRL, Lieb_Hubb_2003_Phys.A}. 
%
%-- HUBBARD DIMER (NM) --
%
\subsection{Hubbard dimer: nonmagnetic solution}
\label{sec:hubbard}
Starting from the half-filled Hubbard dimer, we compare the spectral features of quasiparticle (QP) poles and satellites against the results of Di Sabatino et al.~\cite{Sabatino2021FrontChem} in the case of a self-consistent GW calculation.
The $Z$ factor for a given pole of the GF is $Z_s = \braket{f_s}{f_s} = \text{Tr}[A^s]$, with $A^s$ defined according to Eq.~\eqref{eq:sop_g}.
The exact one-particle GF of the symmetric Hubbard dimer at half filling has two occupied and two empty poles (one quasi-particle and one satellite in each case), symmetric with respect to the Fermi level (here set at zero), because of particle-hole symmetry.
When comparing with the numerically computed GFs (which are in general approximate because of the chosen self-energy),
we consider as QP the pole having the largest $Z$ factor, while we take as satellite the pole having the second largest $Z$ factor.

In this section we limit the study to the nonmagnetic case.
These results are reported in Tab.~\ref{tab:dimer_qp_sat} for both the QP and satellite poles and renormalization factors.
\begin{table}
    \centering
    \begin{tabular}{c c c c c}
    \hline 
    \hline
        $U/t$ & \multicolumn{2}{c}{$\omega_{QP}$} 
              & \multicolumn{2}{c}{$Z_{QP}$}  \\[3pt]
        &  This work & Ref.~\cite{Sabatino2021FrontChem} 
        &  This work & Ref.~\cite{Sabatino2021FrontChem} \\
    \hline 
        1.0   & \textbf{1.0651} & \textbf{1.0651} & \textbf{0.9861} & \textbf{0.9861} \\
        5.0   & \textbf{1.43}26 & \textbf{1.43}34 & \textbf{0.9239} & \textbf{0.9239} \\  
       10.0   & \textbf{1.77}81 & \textbf{1.77}87 & \textbf{0.8777} & \textbf{0.8777} \\ 
       15.0   & \textbf{2.05}25 & \textbf{2.05}42 & \textbf{0.8472} & \textbf{0.8472} \\ 
    \hline 
    \hline
\\[5pt]
    \hline 
    \hline
        $U/t$ & \multicolumn{2}{c}{$\omega_{sat}$} 
              & \multicolumn{2}{c}{$Z_{sat}$}  \\[3pt]
        &  This work & Ref.~\cite{Sabatino2021FrontChem} 
        &  This work & Ref.~\cite{Sabatino2021FrontChem} \\
    \hline 
        1.0   & \textbf{4.0793}  & \textbf{4.0793}  & \textbf{0.0132} & \textbf{0.0132} \\
        5.0   & \textbf{7.63}38  & \textbf{7.63}89  & \textbf{0.0593} & \textbf{0.0593} \\
       10.0   & \textbf{10.8}143 & \textbf{10.8}296 & \textbf{0.082}1 & \textbf{0.082}3 \\
       15.0   & \textbf{13.3}356 & \textbf{13.3}847 & \textbf{0.093}1 & \textbf{0.093}4 \\
    \hline 
    \end{tabular}
    \caption{Hubbard dimer treated at the GW level compared with Ref.~\cite{Sabatino2021FrontChem}. We report the QP (upper panel) and satellite (lower panel) pole positions and their $Z$ factors. The Fermi level is shifted at $E_F=0$. The thresholds used in the calculations are $\delta_M=10^{-2}$ and $\delta_P=10^{-6}$, with a convergence threshold of $\epsilon=10^{-8}$. Frequencies are reported in units of $t$.
    We have highlighted in bold text the digits for which our results agree with those of the reference work.
    \label{tab:dimer_qp_sat}
    }
\end{table}
Overall, we obtain an excellent numerical agreement with the work of Ref.~\cite{Sabatino2021FrontChem}, where a meromorphic representation of the dynamical operators is also employed, with the poles of the GF found via a root finding algorithm.
The agreement in the case of the $Z$ factors is almost perfect, while energy differences around $10^{-3}$ (in units of $t$) are found for the QP poles and energy differences of $10^{-2}$ are found for the satellites in the regime of large $U/t$.
%
%
% -- HUBBARD DIMER (AFM) --
%
\subsection{Hubbard dimer: antiferromagnetic solution}
\label{sec:hubbard_sb}
Within this Section we discuss in detail the features of the broken-symmetry solutions obtained in the Hubbard dimer.
In passing we stress that the exact ground state of a one-dimensional Hubbard chain at half-filling and with an even number of sites is always NM and with a homogeneous charge density~\cite{Lieb_Hubb_1968PRL, Lieb_Hubb_2003_Phys.A} 
, as also found by high-accuracy and exact diagonalization methods~\cite{Martin-Reining-Ceperley2016book,Kotliar2006RevModPhys,Joost2021ContrPlasmaPhys,Honet2022AIP}.  
AFM solutions for the ground state are then unphysical, in the sense that they violate the symmetry of the exact solution. 
Nevertheless, the study of these solutions is still relevant, as it may allow one to obtain better estimates for thermodynamic and spectral quantities, such as the total energy and energy gap.
In fact, following the observation of L\"owdin~\cite{Lowdin_LykosPratt_RevModPhys_1963}, in the case of Hartree-Fock calculations lifting the symmetry constraints on the eigenfunctions may improve the variational flexibility of the method.
The same consideration has been extended to the case of dynamical self-energies such as 2B~\cite{Joost2021ContrPlasmaPhys} and GW~\cite{Honet2022AIP}.

Using our fully self-consistent implementation we investigate the symmetry breaking, specializing to the case of GW.
In order to quantitatively characterize AFM solutions, and following Ref.~\cite{Honet2022AIP}, we introduce the following long-range order parameter:
\begin{equation}
\label{eq:hubbard_afm_order_param}
    M^{LRO} = \sum_i \frac{|n_{i\uparrow} - n_{i\downarrow}|}{2}.
\end{equation}
We observe that in the case of the Hubbard dimer HF gives an AFM ground state for values of $U/t > 2$, similarly to what found in Ref.~\cite{Joost2021ContrPlasmaPhys}.
We also stress that, within HF, no AFM self-consistent solution is actually found for $U/t < 2$, while for $U/t > 2$ we can find either NM or AFM solutions by changing the symmetry of the starting GF of the self-consistency loop.
\begin{figure}
    \centering
    \includegraphics[width=0.45\textwidth]{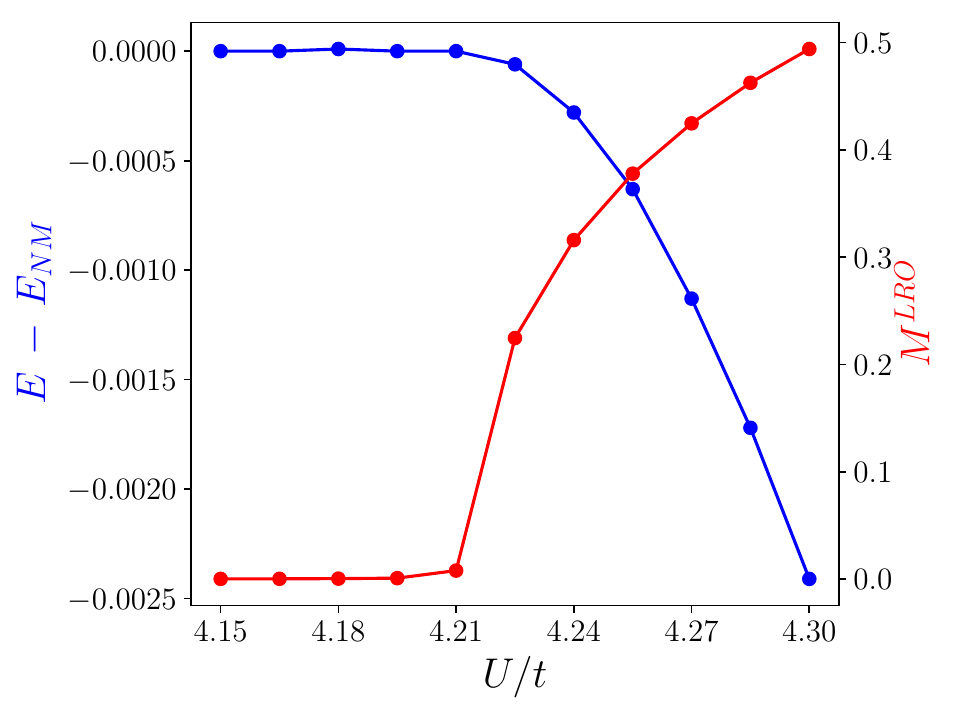}
    \caption{Study of the AFM symmetry breaking in the Hubbard dimer found by GW. We report the difference between the energy $E$ of the unrestricted-symmetry calculation with respect to the non-magnetic (NM) one (left axis). We also report the order parameter evaluated for the unrestricted symmetry solution (right axis).}
    \label{fig:hub_ssb_GW}
\end{figure}

The GW calculations are initialized from the HF self-consistent solutions obtained at the corresponding value of $U/t$.
We observe that, at variance with Ref.~\cite{Honet2022AIP}, even if the self-consistency loop starts from an AFM GF, it does not necessarily converge to an AFM solution,
while a NM starting GF always restricts the symmetry of the ground state.
These results are reported in Fig.~\ref{fig:hub_ssb_GW}, where we study the energy difference (left axis) between the unrestricted symmetry (US) calculation with respect to the NM one.
In Fig.~\ref{fig:hub_ssb_GW} we also report the order parameter (right axis) of the self-consistent GF at the end of the loop for the US case.
In the case of GW we find that an AFM ground state is found for $U/t > 4.21$, while there is no AFM solution below that value, signaled by the vanishing order parameter $M^{LRO}$ in Fig.~\ref{fig:hub_ssb_GW}.
We notice that the AIM-SOP implementation is capable of high accuracy in resolving total energy differences also below $10^{-3}$ (units of $t$ are used throughout). 
Furthermore the presence of an AFM solution at $U/t\simeq 4.21$ is signaled by the simultaneous onset of the order parameter $M^{LRO}$ at the same value of $U/t$ where the energy difference $E-E_{NM}$ becomes negative. The AFM solution persists also in the atomic limit, where it gives rise to a finite energy gap that scales with $U$, consistently with the exact result and at variance with the metallic NM solution within GW~\cite{Lieb_Hubb_1968PRL, correl21_book, DiSabatino_PhysRevB.94.155141_2016}. 
\begin{figure*}
    \centering
    \includegraphics[width=1.0\textwidth]{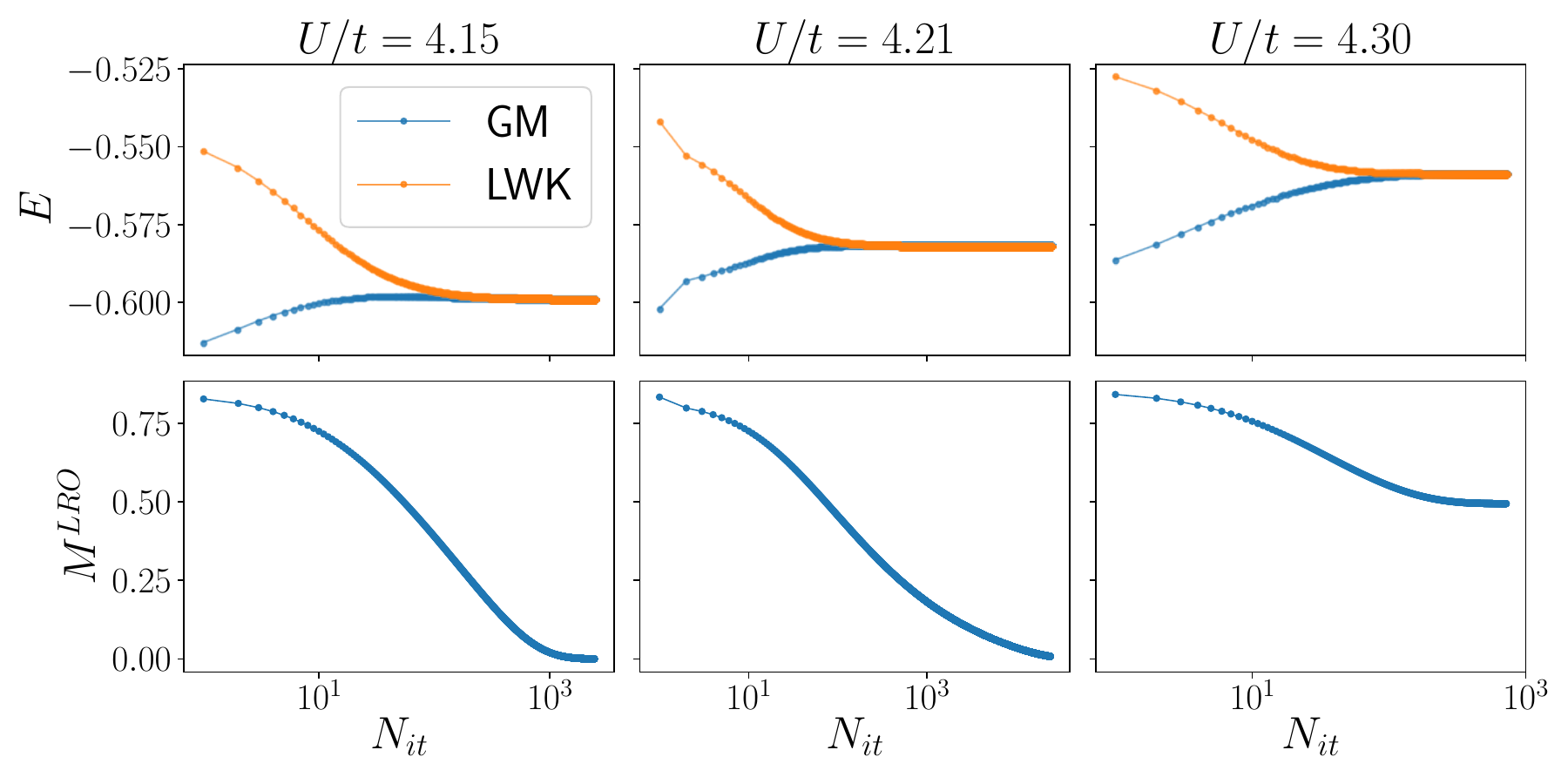}
    \caption{Total energies (first row) and values of the order parameter $M^{LRO}$ (second row) during self-consistency convergence, for the half-filled Hubbard dimer studied with GW. $N_{it}$ is the index of the iteration step. The total energy is calculated both with the Galitskii-Migdal (GM) formula and the Klein (LWK) functional. 
    The number of iterations is reported on a logarithmic scale.}
    \label{fig:hub_ssb_GW_run}
\end{figure*}

In order to assess the stability of our AIM-SOP implementation, we study how the total energy and the order parameter change within the self-consistency loop.
In Fig.~\ref{fig:hub_ssb_GW_run} we report the values of the total energy and of the order parameter for the half-filled Hubbard dimer obtained in GW for an US calculation. 
We observe that when the stopping condition of the loop is reached, the Klein functional of Eq.~\eqref{eq:klein_functional} agrees with the energy value returned from the Galitskii-Migdal formula, as expected.
In general the agreement between different functionals (or with the GM formula) is a non-trivial result and it can also be taken as a diagnostic that self-consistency has been achieved~\cite{Friesen2010PhysRevB}.

We note that for $U/t < 4.21$ the loop converges towards a NM solution, by slowly reducing the order parameter at each iteration.
Instead, above the value of $U/t>4.21$ the loop converges to an AFM self-consistent solution.
The order parameter is finite in this situation, but it is smaller than that of the corresponding HF AFM solution.
Another relevant consideration in this case is that the total energy reaches a stable value earlier with respect to the order parameter, so that relative differences in the order parameter for this system may be more evident than differences in total energies.
Overall, the total energy and the order parameter evolve smoothly during the self-consistency loop, assuring the stable behaviour of our AIM-SOP implementation.
At variance to GW, when using a 2B self-energy AFM solutions can be stabilized even when they are not the ground state of the system.
Broken symmetry solutions of the Hubbard dimer found using the 2B are discussed in detail in the Supplemental Material~\cite{suppinfo}, where we also comment on the difference with respect to GW.

We stress that within our AIM-SOP implementation, whenever starting from a non-interacting $G_0$ having real poles and positive semi-definite (PSD) residues, we only manage to find at most one self-consistent solution for each symmetry (NM or AFM) at any given value of $U/t$.
Thus multiple solutions with the same symmetry have not been found, neither considering different starting points as discussed for the 2B self-energy in the Supplemental Material~\cite{suppinfo}, nor adopting different mixing schemes as discussed in Sec.~\ref{sec:self_consistency}.
Furthermore, we remark that the poles of the GF at each step of the self-consistency loop are always found on the real axis if the starting GF has poles on the real axis, and the residues of the GF are always PSD for the self-energy approximations here considered. 
This implies that at each step of the loop the GF is compatible with a Lehmann representation~\cite{Lehmann_NuovoCimento_1954}, and that all the expected analytical properties are satisfied~\cite{Ferretti2024PRB}.
In particular, as analyzed in Ref.~\cite{Ferretti2024PRB}, if the input self-energy is well-behaved, i.e.\ if it has Hermitian PSD residues, real poles, and Hermitian asymptotic value $\Sigma_0$, the AIM-SOP procedure leads to a Hermitian effective AIM Hamiltonian, out of which the resulting GF is also well-behaved (PSD residues and real poles). The reverse is also true~\cite{Ferretti2024PRB}.
%
%
% -- SPECTRAL PROPERTIES --
%
\subsection{Hubbard chains: spectral properties}
\label{subsec:spectral_hubbard_chains}
We now move to study the spectral features of longer Hubbard chains, where the QP and satellite structure is richer with respect to that of the Hubbard dimer.
The results of our AIM-SOP implementation are compared with those of Refs.~\cite{Friesen2010PhysRevB,Joost2021ContrPlasmaPhys}.
We start by considering a $N=6$ Hubbard chain at half-filling, with $U/t = 4$ and open boundary conditions, and focus on the NM solution. 
\begin{figure}
    \centering
    \includegraphics[width=0.45\textwidth]{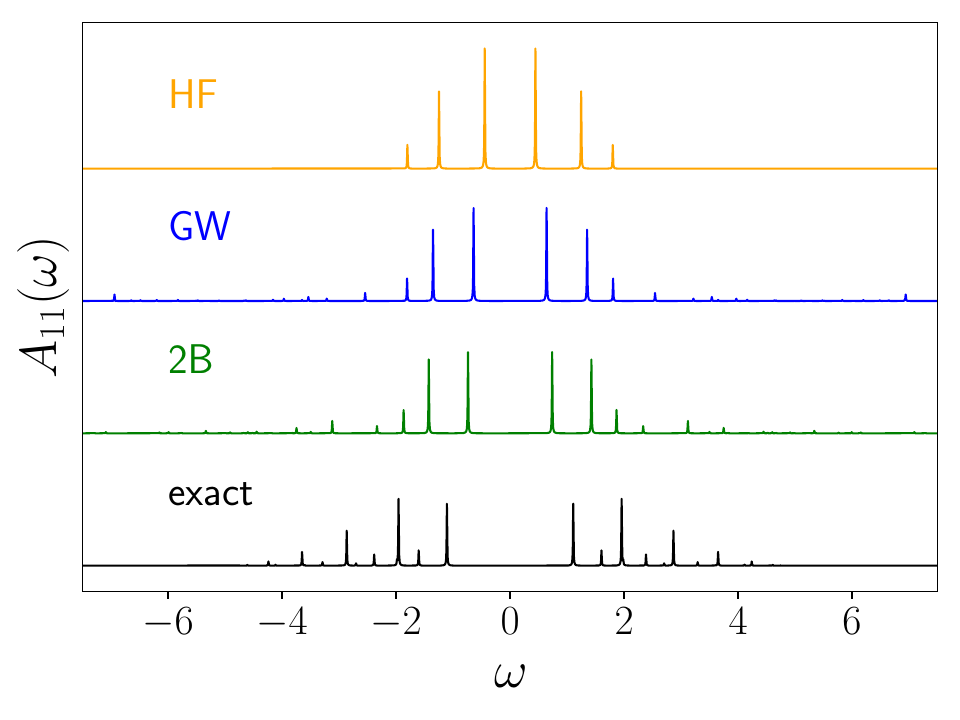}
    \caption{Spectral density $A_{11}(\omega)$ on the first site of an open Hubbard chain with $N=6$ and $U/t=4$. The value of $\omega$ is scaled with the hopping term $t$ and the Fermi level has been shifted to $\omega=0$, while $A_{11}$ is reported in arbitrary units. The spectral function is reported for different self-energy approximations and for the NM ground state. The exact solution is also reported. The plot has been obtained by including a finite imaginary shift $\eta=5\times 10^{-3}$ in the poles of the GF.}
    \label{fig:spec_N6_U4_comp_friesen}
\end{figure}
The spectral density projected on the first site of the chain is reported in Fig.~\ref{fig:spec_N6_U4_comp_friesen}.
We show the spectral density obtained with the different approximations and also the exact result obtained via exact diagonalization.
Overall we obtain an excellent agreement with the work of von Friesen et al.~\cite{Friesen2010PhysRevB} for the HF, 2B, and GW levels of theory.
In particular, HF gives the same spectral density of the non-interacting system, consisting of three main occupied peaks (plus particle-hole symmetry). 
Besides this energy region, the 2B self-energy gives satellites with larger spectral weight,
while in GW a structureless spectrum of satellites appears. 

As also stressed in Ref.~\cite{Friesen2010PhysRevB}, we notice that the self-consistency of GW in longer chains and with large $U/t$ becomes slow and difficult, since this self-energy approximation gives rise to several satellites with small residues.
In particular, for GW the position of the furthest satellite from the Fermi level (around $\omega/t=7.5$) changes weakly with the pruning threshold. 
The residue of this satellite is also larger with respect to the nearby ones since, due to our pruning scheme, it is receiving the spectral weight of the pruned poles far away from the Fermi level. 

\begin{figure}
    \centering
    \includegraphics[width=0.45\textwidth]{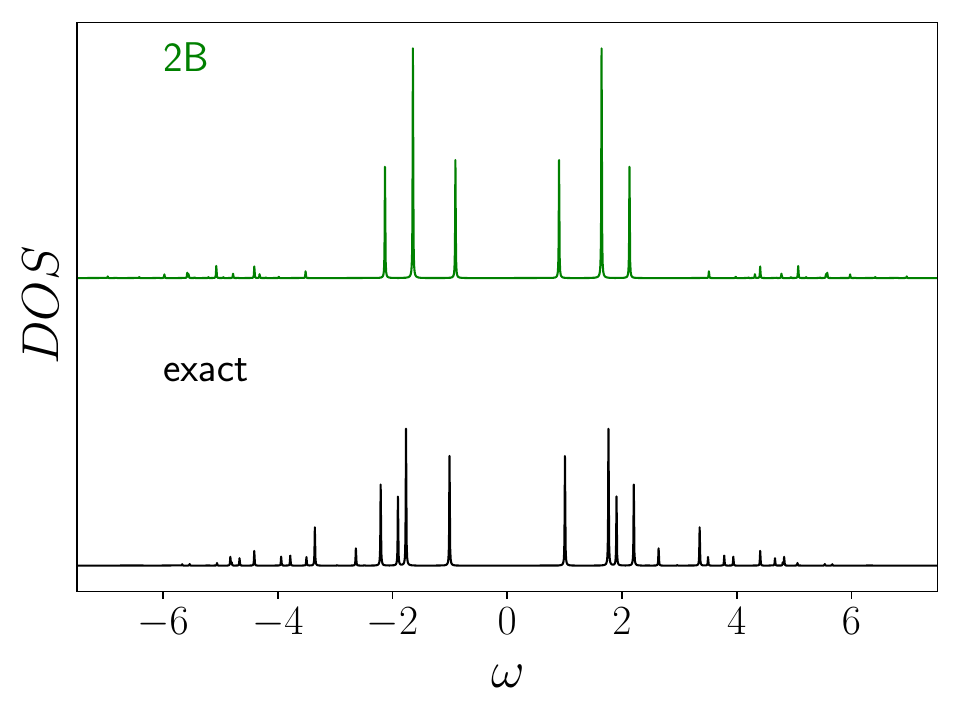}
    \caption{Density of states (DOS) of the $N=8$ Hubbard chain with periodic boundary conditions and $U/t=4$ for an AFM ground state found within the 2B self-energy. The value of $\omega$ is scaled with the hopping parameter $t$ and the chemical potential has been shifted to $\omega=0$, while the DOS is reported in arbitrary units. The plot has been obtained by adding a finite imaginary shift $\eta=5\times 10^{-3}$ to the poles of the GF.}
    \label{fig:hub_N8_U4_spectral}
\end{figure}

We then consider the case of a $N=8$ Hubbard chain with periodic boundary conditions and $U/t=4$.  
The same system has been considered by Joost et al.\ in Ref.~\cite{Joost2021ContrPlasmaPhys}, where a real-time implementation has been employed.
We focus here on an AFM solution found with the self-consistent 2B self-energy.
The density of states (DOS) of this chain is reported in Fig.~\ref{fig:hub_N8_U4_spectral}.
We find an excellent agreement with respect to the DOS of Ref.~\cite{Joost2021ContrPlasmaPhys}, while we also observe a very good agreement for the value of the total energy: in our case we obtain $E^{LWK}_\text{tot}/t = -4.069(0)$ and $E^{GM}_\text{tot}/t = -4.069(3)$, against $E^\text{ref}_\text{tot}/t = -4.08$.
We notice how the AIM-SOP implementation is also able to resolve well the presence of the satellites for the 2B approximation between $\omega = -6.0 t$ and $\omega = -4.0 t$.
In passing, we note that we could not exactly reproduce the results of Ref.~\cite{Honet2022AIP}. 
We believe that a different self-consistency condition may be among the possible causes of discrepancy.
A detailed study of a $N=6$ Hubbard chain, both in the NM and unrestricted-symmetry calculations, is reported in the Supplemental Material~\cite{suppinfo}.
%
%
%========= AFM & CDW WITH LR INTERACTION ========
%
%
\section{Long-range interaction}
\label{sec:long-range}
In this Section we consider a lattice model including a long-range interaction term, extending the Hubbard Hamiltonian of Eq.~\eqref{eq:hubbard_hamiltonian}.
Similarly to what done in Sec.~\ref{sec:validation}, we restrict to one orbital per site and to systems at half-filling in periodic boundary conditions, so that the model is particle-hole symmetric.
The Hamiltonian is given by
\begin{align}
    \label{eq:long-range_ham}
    H = &- t \sum_\sigma \sum_{\langle i,j \rangle} 
    \left(c^\dagger_{i\sigma } c_{j\sigma} + h.c.\right)
    + U\sum_i n_{i\uparrow }n_{i\downarrow} \\
    \nonumber 
    &+ \frac{1}{2}\sum_{\sigma,\sigma'}\sum_{i\neq j}V_{ij}n_{i\sigma }n_{j\sigma'},
\end{align}
where the long-range interaction between site $i$ and $j$ is determined by
\begin{equation}
    \label{eq:long-range_Vint}
    V_{ij} = \frac{V}{R_{ij}},
\end{equation}
with $V$ being a parameter that defines the energy scale of the non-local interaction. In the above expression, $R_{ij}$ is the distance between the two sites (in units of the nearest-neighbor distance).

We note in passing that different parametrizations have also been employed in similar models accounting for a non-local electron-electron repulsion~\cite{Ohno_TheoreticaChimicaActa_1964,Kaasbjerg_GWmodels_PhysRevB_2010,Strange_PhysRevB_2012}. 
All these parametrizations, however, share the same $V_{ij} \propto 1/R_{ij}$ behavior at large distances.  
The model depends on two dimensionless parameters: the ratio $U/t$ defines the strength of the correlation, while the ratio $V/U$ sets the strength of the non-local interaction term with respect to the local one. 
We take $V/U\leq 1$ as we expect inter-site electron repulsion to be always smaller than the on-site one.

Imposing periodic boundary conditions, the ground state having the same spatial symmetry of the non-interacting system is given by a NM state with a homogeneous charge density.
This allows us to consider two possible broken symmetry solutions: AFM solutions giving an overall homogeneous charge density and charge density waves (CDW) giving a NM solution with an oscillatory charge density.
For both broken symmetry solutions we only consider states where the charge density has a spatial periodicity of two lattice sites.
It is also important to stress that, since an exact numerical study of the system always gives a homogeneous and nonmagnetic ground state at half-filling, the broken symmetry ground state solutions found via MBPT are artifacts of the approximation. 
%
%
%-- RESULTS --
\subsection{Results}
\label{subsec:results}
We analyze the results obtained for a chain with $N=4$ sites. 
Here we fix the value of $U/t$ and we consider how the total energy changes as a function of the ratio $V/U$.
For each value of the parameters considered we are able to converge to the symmetric solution, which has the same symmetry of the non-interacting system, to an AFM solution or to a CDW one.
This is accomplished by changing the symmetry of the starting GF at the beginning of the self-consistency loop, similarly to what is done in Sec.~\ref{sec:hubbard_sb}.
The convergence to the three solutions with different symmetries is achieved for all the self-energy approximations considered.
\begin{figure}
    \centering
    \includegraphics[width=0.45\textwidth]{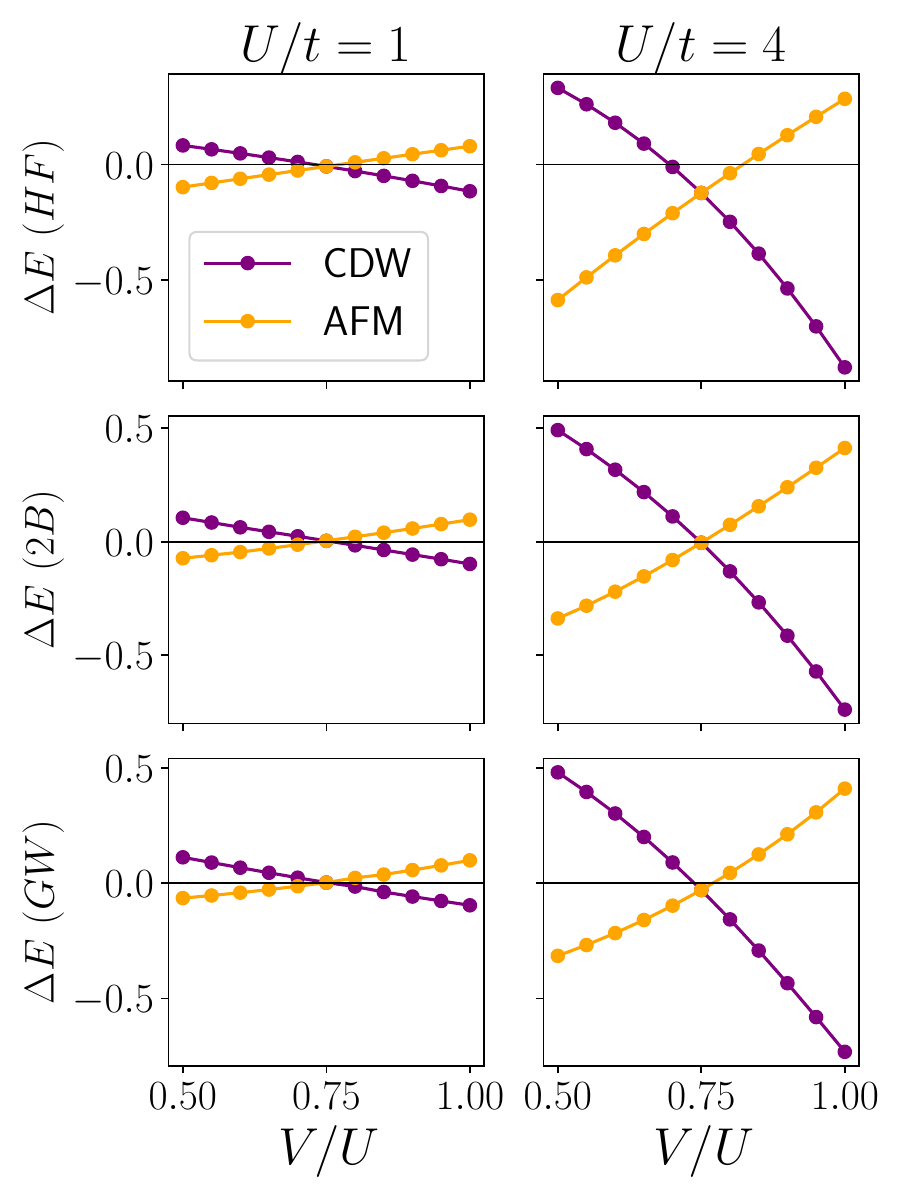}
    \caption{Energy difference between the broken symmetry solutions and the symmetric solution for different values of the model parameters and for different self-energy approximations. 
    The results are reported for a closed chain with $N=4$ sites.
    A cross in energy between the CDW and AFM solutions is observed at $V/U=0.75$ for all the self-energy approximations considered, independently of the value of $U/t$.}
    \label{fig:ediff_all_N4}
\end{figure}

In Fig.~\ref{fig:ediff_all_N4} we report the energy difference $\Delta E = E_{broken} - E_{symm}$, between the 
broken symmetry and the 
symmetric solutions obtained for all the different SE approximations.
We consider values of $U/t = 1$ (weak correlation) and of $U/t=4$ (strong correlation). 
It is interesting to observe that, independently of the value of $U/t$, all the SE approximations predict a cross in energy between the AFM and CDW solutions for $V/U = 0.75$.
The stability of the symmetric solution with respect to the broken symmetry ones changes instead with the value of $U/t$. 

\begin{figure}
    \centering
    \includegraphics[width=0.45\textwidth]{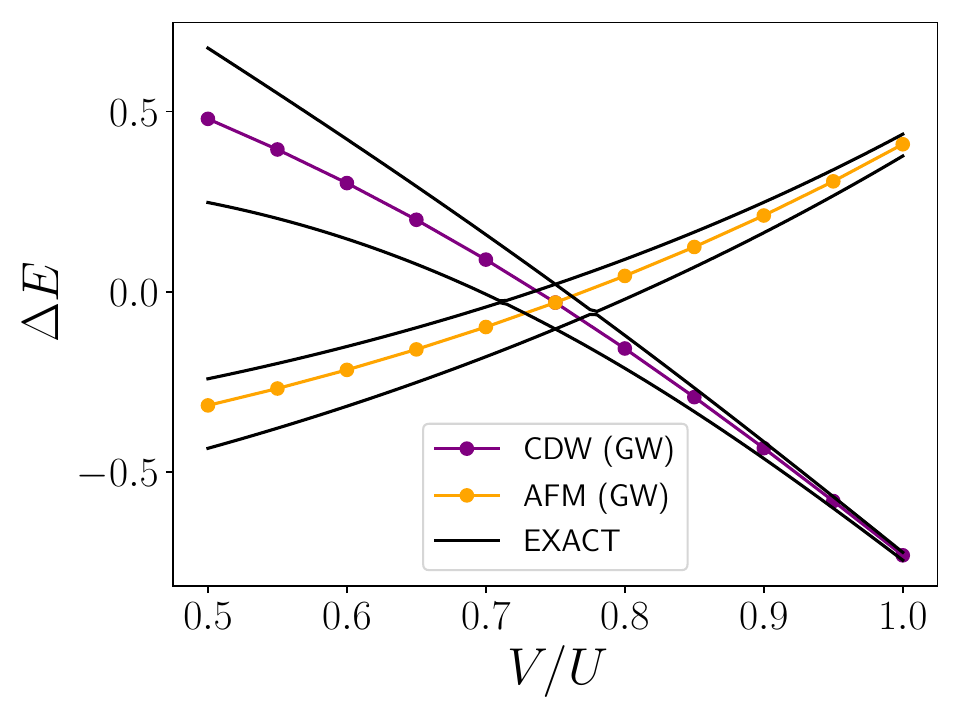}
    \caption{Energy difference between the first exact states of the many-body Hamiltonian and the symmetric solutions found with the GW approximation on a chain with $N=4$ sites. The values of the broken symmetry solutions are also reported.}
    \label{fig:ediff_exact_N4}
\end{figure}

In order to understand this behavior it is instructive to study the exact ground state and the first excited states of the neutral system for the $N=4$ chain, which can be determined through exact diagonalization of the many-body Hamiltonian.
In Fig.~\ref{fig:ediff_exact_N4} we report the energy difference $\Delta E = E_{exact} - E^{GW}_{symm}$ between the first four exact eigenenergies of the neutral many-body Hamiltonian and the total energy of the symmetric state obtained via the GW approximation.
The first excited state at lower values of $V/U$ is actually a spin singlet and it is three-fold degenerate in case of spin unrestricted calculations, similarly to what happens for the Hubbard model.
The exact diagonalization always returns a NM interacting ground state having the same symmetry of the non-interacting ground state. 
It is interesting to notice however that the eigenenergies of the neutral many-body Hamiltonian have a degeneracy at the same value of $V/U=0.75$ observed in Fig.~\ref{fig:ediff_all_N4}.
If the MBPT calculation is allowed to break the symmetry, the prediction of the total energy improves with respect to the symmetric state, and the curve of the total energy obtained via approximate MBPT follows the exact one.

The nature of the exact ground state can be further characterized by the introduction of appropriate correlation functions involving two density operators. 
In particular, even if the exact ground state shares the same symmetry of the non-interacting one, it can still develop an AFM or CDW character, which can be measured by the evaluation of ad hoc correlation functions.
This is, for instance, the case of the Hubbard model, whose ground state develops strong AFM correlations in the large $U/t$ regime without breaking explicitly the symmetry of the Hamiltonian~\cite{Joost2021ContrPlasmaPhys,Honet2022AIP}.

The AFM character can be evaluated by considering the following correlation function
\begin{equation}
\label{eq:corr_afm}
     C^{AFM} := \frac{1}{N}\sum_{i=1}^{N} \langle n_{i\uparrow} n_{i\downarrow} \rangle ,
\end{equation}
which measures the double occupancy for spin-up and spin-down electrons to occupy the same site.
Instead, the CDW character could be related to the following 
\begin{equation}
\label{eq:corr_cdw}
    C^{CDW} := \frac{1}{4}\frac{1}{N}\sum_{i=1}^{N}
                    \sum_{\sigma\sigma'} \langle  n_{i-1\sigma} n_{i\sigma'} \rangle ,
\end{equation}
which measures the overall charge dimerization on two adjacent sites.
The factor of $1/4$ in Eq.~\eqref{eq:corr_cdw} takes into account the number of all spin combinations entering the sum and makes Eq.~\eqref{eq:corr_cdw} comparable to Eq.~\eqref{eq:corr_afm}.
The smaller the value of the correlation functions in Eqs.~\eqref{eq:corr_afm} and~\eqref{eq:corr_cdw}, the larger the corresponding AFM or CDW character.
While these correlation functions are easily accessible in the exact calculations, as they can be directly computed through the action of field operators on the many-body wavefunctions, they can not be directly evaluated within the present MBPT approach, where we have mostly access to the one-particle Green's function ($C^{AFM}$ in Eq.~\eqref{eq:corr_afm} is easily accessible only in the Hubbard model, where the local nature of the electron-electron interaction allows one to extract this correlation function from the knowledge of the total and interaction energies~\cite{Joost2021ContrPlasmaPhys,Honet2022AIP}). 
For the broken symmetry solutions obtained within approximate MBPT calculations, we then consider a ``factorized" version of the former correlation functions:
\begin{align}
\label{eq:corr_afm_mbpt}
    \widetilde{C}^{AFM} &= \frac{1}{N} \sum_{i=1}^{N} \langle n_{i\uparrow} \rangle \langle n_{i\downarrow} \rangle, \\
\label{eq:corr_cdw_mbpt}
    \widetilde{C}^{CDW} &= \frac{1}{4}\frac{1}{N}\sum_{i=1}^{N}
                    \left( \sum_{\sigma} \langle  n_{i-1\sigma} \rangle \right) \left(\sum_{\sigma'} \langle n_{i\sigma'} \rangle \right).
\end{align}

\begin{figure}
    \centering
    \includegraphics[width=0.45\textwidth]{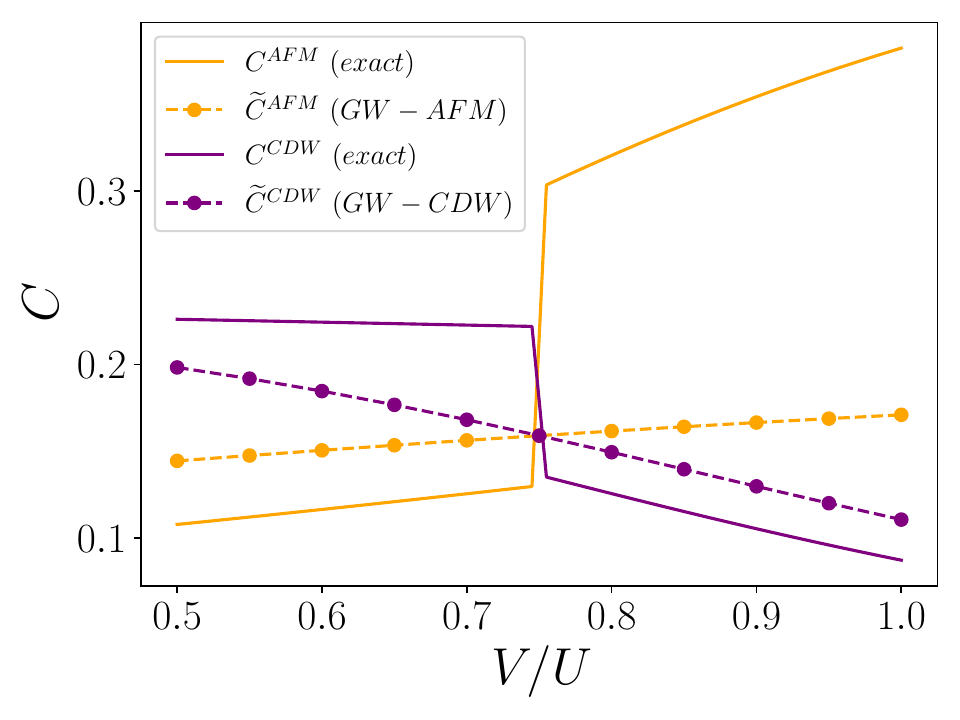}
    \caption{Correlation functions evaluated on the ground state of the $N=4$ chain with $U/t=4$.
    The solid line gives the $C^{AFM}$ and $C^{CDW}$ correlation functions evaluated for the exact ground state, while the dashed lines reports the ``factorized" $\widetilde{C}^{AFM}$ and $\widetilde{C}^{CDW}$ correlation functions evaluated for the GW solution with the corresponding symmetry.}
    \label{fig:corr_all}
\end{figure}

In Fig.~\ref{fig:corr_all} we report the correlation functions in Eqs.~\eqref{eq:corr_afm} and~\eqref{eq:corr_cdw} for the exact ground state (solid lines) and the value of the ``factorized" correlation functions in Eqs.~\eqref{eq:corr_afm_mbpt} and~\eqref{eq:corr_cdw_mbpt} for the GW solutions (dashed line) in the case of the corresponding broken symmetry solutions. 
The model parameters are $U/t=4$ and the parameter $V/U$ takes values in the interval $[0.5,1.0]$.
We observe that at the value of $V/U=0.75$, for which the first two eigenstates of the many-body Hamiltonian are degenerate, the correlation functions change in a discontinuous way, signaling a change of character of the exact ground state.
Remarkably, while there is no agreement in absolute values between the GW results and the exact ones, the symmetry breaking in the GW calculation is able to correctly describe the trend of how the correlation functions change with respect to the parameter $V/U$.

Importantly, for this particular system the symmetry breaking at the MBPT level, while unphysical, follows from the development of analogous correlations in the exact system.
Interpreting the ratio $V/U$ in terms of the distance between the sites of the chain -- i.e. a small value of $V/U$ representing a large inter-site distance we can see that AFM correlations dominate in the large distance regime as observed for hydrogen chains~\cite{Motta_PRX_2020}, while CDW correlations are relevant at smaller inter-site distances~\cite{Barborini_PRB_2022}. 
%
%
%-- SOLUTION OF SSE --
%
\subsection{Solutions of the SSE}
\label{subsec:delta_xc_long_range}
In this section we analyze how the broken symmetry solutions studied in Sec.~\ref{subsec:results} are translated to the KS-DFT level of description.
In order to do so we focus on a chain with periodic boundary conditions and $N=16$ sites. 
The local electron-electron repulsion is fixed at $U/t=4$ and self-consistent calculations are performed with GW.
We observe that the broken symmetry solutions with AFM and CDW symmetry can also be obtained through the Sham-Schl\"uter equation (SSE) presented in Sec.~\ref{subsec:sse_theory}, and so they are also present in the Kohn-Sham system.

It is interesting to stress that, due to the simplified nature of the lattice model, the symmetric solution leads to a uniform charge density, which is then the same for all the values of the model parameters.
This means that, for the symmetric solutions, the KS potential $v^\text{KS}$ is just a constant term that aligns the highest occupied molecular (quasiparticle) orbital (HOMO) of the non-interacting ground state to the HOMO of the interacting symmetric state.
In particular, in the case of chains having a metallic non-interacting ground state (see Supplemental Material~\cite{suppinfo}), the Kohn-Sham Green's function $G^\text{KS}$ describes a metallic system, while the system displays a gap at the MBPT level.
We stress here that in order to converge the GW calculations in the case of the symmetric state, the self-consistency loop is initialized with a gapped GF.

\begin{figure}
    \centering
    \includegraphics[width=0.45\textwidth]{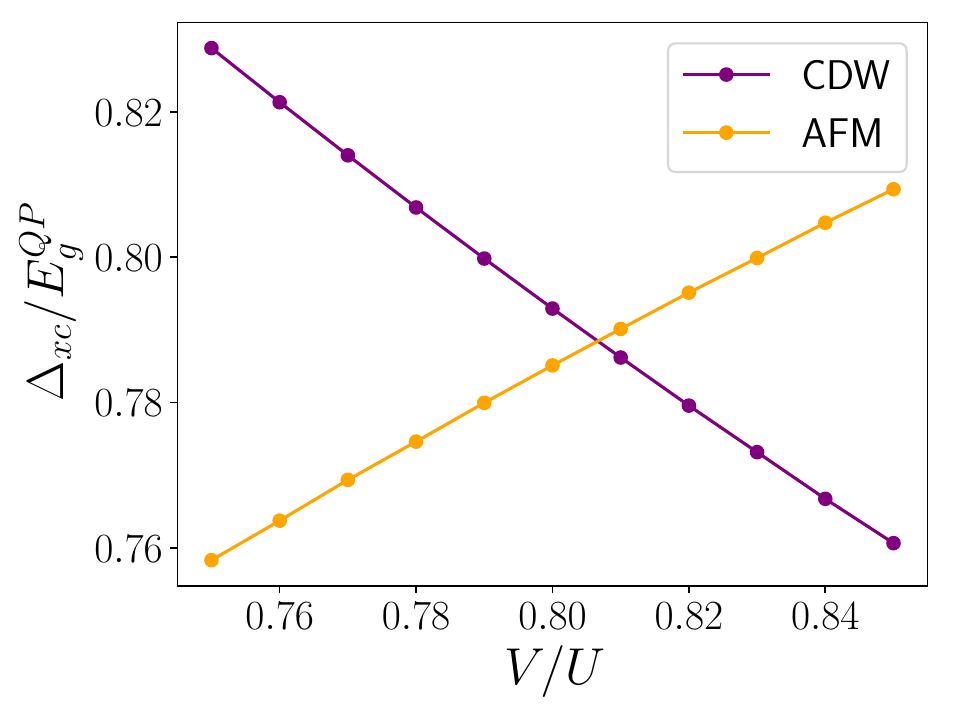}
    \caption{Ratio between the derivative discontinuity of the Kohn-Sham potential $\Delta_{xc}$ and the quasiparticle energy gap $E_g^{QP}$ in the case of GW and for a periodic chain of $N=16$ sites with $U/t=4$. The derivative discontinuity is evaluated via Eq.~\eqref{eq:delta_xc} and we report only the case of AFM and CDW solutions.}
    \label{fig:delta_xc}
\end{figure}

Instead, in the case of broken symmetry solutions, $G^\text{KS}$ describes a gapped system at finite values of $U$ and $V$.
The ratio between the derivative discontinuity $\Delta_{xc}$ and the quasiparticle energy gap is reported in Fig.~\ref{fig:delta_xc}.
We observe that the derivative discontinuity for the $N=16$ chain is large and accounts for $70\%$ to $80\%$ of the quasiparticle energy gap. 
Similar results have also been observed in model 1D semiconductors with long-range electronic interactions~\cite{SchonhammerGunnarsson_JournalOfPhysicsC_1987}.
As an interesting point, the ratio between $\Delta_{xc}$ and $E_g^{QP}$ gets smaller by increasing the parameter $V/U$ for the CDW solution, while the opposite happens for the AFM solution.
For the $N=16$ chain the value of $V/U$ at which the two curves of Fig.~\ref{fig:delta_xc} cross  is also the value of $V/U$ at which the energy of the AFM solution becomes larger than that of the CDW one at the level of GW.
Having a large derivative discontinuity, the energy gap of $G_{KS}$ in the case of broken symmetry solutions is severely underestimated with respect to the quasiparticle gap.

\begin{figure}[]
    \centering
    \includegraphics[width=0.45\textwidth]{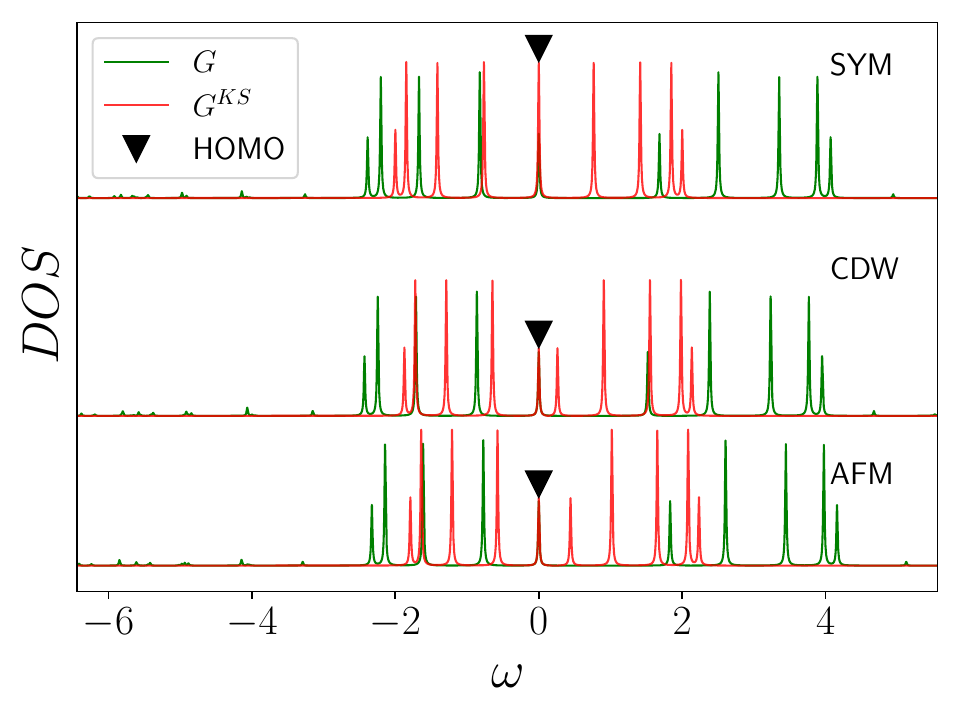}
    \caption{DOS of the $N=16$ chain with periodic boundary conditions at $U/t=4$ and $V/U=0.5$ for a long-range Coulomb interaction.
    We report the DOS for all states with different symmetry.
    The green line is related to the GW DOS, while the red line is the DOS of the corresponding non-interacting $G^\text{KS}$.
    The HOMO quasiparticle level is marked with a downwards triangle and shifted to $\omega=0$ in all cases.
    The frequency is scaled with the hopping parameter $t$ and the DOS is reported in arbitrary units.}
    \label{fig:spectral_Gks}
\end{figure}

In Fig.~\ref{fig:spectral_Gks} we report the DOS of the periodic $N=16$ chain at $U/t=4$ and $V/U=0.50$.
The DOS has been evaluated in the case of the GW for the symmetric as well as broken symmetry solutions.
The green line reports the DOS of the self-consistent GF, while the red line reports the DOS of $G^\text{KS}$.
The position of the HOMO level is also highlighted. 
As discussed in Sec.~\ref{sec:self_consistency}, the system of linear equations defined by the SSE, Eq.~\eqref{eq:sse}, is solved requiring that the HOMO level of $G^\text{KS}$ is aligned to that of $G$.
As stated before, in the case of the symmetric solution, $G_{KS}$ describes a metallic system instead of a gapped one.
The broken symmetry KS GF's are instead gapped, but as a consequence of the large derivative discontinuity the KS gap is severely underestimated with respect to the GW quasiparticle gap.

In conclusion, the unphysical broken symmetry solutions found via approximate self-energies at the MBPT level can be inherited at the KS-DFT level.
Furthermore, for our lattice model (1D systems having a metallic ground state) the derivative discontinuity in the KS exchange-correlation potential can be extremely large.
In this case we also observe that the occupied KS energy levels do not describe accurately the one-particle excitation energies of the system.
%
%
%========= CONCLUSIONS ==========================
%
\section{Conclusions}
\label{sec:conclusions}
In this work we study one-dimensional lattice models with an approximate many-body perturbation theory (MBPT) treatment of the electron-electron interaction.
The Dyson equation is solved in a fully self-consistent manner via the algorithmic-inversion method based on the sum-over-pole representation of dynamical operators (AIM-SOP)~\cite{Chiarotti2022PRR,Chiarotti2023PhD,Chiarotti2024PRR}.
Our calculations demonstrate that the AIM-SOP framework enables to solve self-consistently the Dyson equation in a stable and accurate way, for the system properties considered, obtaining an excellent agreement with the existing literature~\cite{Friesen2010PhysRevB,Sabatino2021FrontChem,Joost2021ContrPlasmaPhys}.
As a result, the fully self-consistent implementation allows one to study in detail the total energy of broken symmetry solutions by evaluating unambiguously the value of the Klein functional at self-consistency. 
Notably, in this work we make use of -- and we further numerically validate -- a recent analytical expression to compute the term $\TrLn\{G_0^{-1}G\}$ appearing in the Klein functional~\cite{Ferretti2024PRB}.

We then apply the AIM-SOP framework to study broken symmetry solutions, extending the results reported in Refs.~\cite{Joost2021ContrPlasmaPhys, Honet2022AIP}, to  model systems accounting also for non-local electronic interactions.
We show that -- while the exact ground state has  a homogeneous charge density as in the case of the Hubbard model -- broken symmetry solutions with a charge density wave (CDW) or antiferromagnetic (AFM) ordering can be found within an approximate MBPT treatment.
We consider the example of a 4-site chain where the unphysical MBPT broken symmetry solutions can be traced back to the development of an AFM or CDW character in the exact ground state, which is related to the interplay between the local and the non-local electronic interactions and can be probed by density-density correlation functions such as those of Eq.~\eqref{eq:corr_afm} and Eq.~\eqref{eq:corr_cdw}.

In particular, the CDW character gets enhanced when the strength of the  non-local electronic interaction $V$ approaches the strength of the local electronic repulsion $U$.
For the 4-site chain we show that the ground state switches from AFM to CDW when the ratio between the local and non-local interaction of Eq.~\eqref{eq:long-range_Vint} becomes larger than $V/U=0.75$, independently of the value of $U/t$.
Moreover, the estimation of the total energy with the GW approximation improves considerably for symmetry unrestricted calculations, both for AFM and CDW solutions.
We also observe that broken symmetry solutions in longer chains are found when the non-interacting ground state of the model Hamiltonian in Eq.~\eqref{eq:long-range_ham} is metallic.

Finally, we study how the broken symmetry solutions are translated in the framework of Kohn-Sham density functional theory (KS-DFT) by adopting the AIM-SOP framework to solve the (non-linearized) Sham-Schl\"uter equation. 
We verify that in this model, if the non-interacting ground state is metallic, the KS solution is always metallic in case of symmetry restricted calculations, while a small gap opens at the KS level of theory for broken symmetry solutions.
Furthermore, we show for a 16-site chain that if the model of Eq.~\eqref{eq:long-range_ham} has a non-interacting metallic ground state, then the derivative discontinuity in the corresponding KS potential tends to be remarkably large, further stressing 
the relevance of this exchange-correlation driven correction to the KS gap.
\bigskip
%
% ======== ACKNOWLEDGEMENTS =====================
%
\begin{acknowledgments}
We acknowledge insightful discussions with N.~Marzari, D.~Varsano, G.~Lani and A.~Pintus.
T.C. acknowledges the support from the Swiss National Science Foundation (SNSF) through Grant No. 200020\_213082 for this work.

\end{acknowledgments}
%
% ======== APPENDIX =============================
%
\bigskip
\appendix
%
% -- SOP SELF-ENERGY --
\section{SOP representation of the HF, GW, and 2B self-energies}
\label{app:sop_self_ene}
In this Appendix we report the SOP representation of the self-energy approximations used throughout our work, namely HF, GW, and 2B. Throughout this section, repeated indexes $i$, $j$ and $k$ are meant to be summed over. 
Since Green's function and self-energy operators are diagonal in their spin indexes we do not write them explicitly, unless otherwise stated.
The Green's function of the interacting system $G(\omega)$, the RPA polarizability $P(\omega)$, and the screened interaction $W(\omega)$ are defined by the following convention about poles and residues:
\begin{align}
    G_{ij}(\omega) &= \sum_s \frac{A^s_{ij}}{\omega - z_s},\\
    P_{ij}(\omega) &= \sum_t \frac{R^t_{ij}}{\omega - \Omega_t},\\
    W_{ij}(\omega) &= \sum_t \frac{B^t_{ij}}{\omega - w_t},
\end{align}
where the indexes $i,j$ refers to 
%the matrix elements of the operator in the basis of 
the lattice coordinates.
The density-matrix $\gamma_{ij}$ is related to $G(\omega)$ by:
\begin{equation}
    \gamma_{ij} = \Trw[G_{ij}(\omega)] = \sum_s^{occ} A^s_{ij}, 
\end{equation}
where the ``occupied" label indicates poles with energies below the Fermi level. 
The Hartree-Fock self-energy is then given by
\begin{equation}
    \Sigma^{HF}_{ij} = \delta_{ij} n_{k} v_{ki} - \gamma_{ij} v_{ji} ,
\end{equation}
where $n_k$ is the total charge density on site $k$, $\gamma_{ij}$ is the spin-resolved density matrix, and $v_{ij}$ are the matrix elements of the electron-electron repulsion term in the basis of the lattice coordinates.

The 2B self-energy is obtained by adding two second order terms (direct and exchange) to the HF one. 
%The SOP expression of the second order direct term is:
\begin{widetext}
\begin{eqnarray}
%\nonumber
\label{eq:2bd_sgm}
    \Sigma^{2B,d}_{ij}(\omega) &=& - \int \frac{d\omega_1}{2\pi i}
    %e^{i\omega_1 0^+} 
    G_{ij}(\omega + \omega_1) \, v_{ik} \, P_{kl}(\omega_1)v_{lj} %= \\
    = \sum_s^{occ}\sum_t^{emp} \frac{A^s_{ij}v_{ik} R^t_{kl} v_{lj} }{\omega - (z_s - \Omega_t)} - \sum_s^{emp}\sum_t^{occ} \frac{A^s_{ij}v_{ik} R^t_{kl} v_{lj} }{\omega - (z_s - \Omega_t)}.
%\end{eqnarray}
%Instead, the second order exchange term is given by:
%\begin{align}
\\[7pt]
    \label{eq:2bx_sgm}
    \Sigma^{2B,x}_{ij}(\omega) &=& \phantom{-}\int \frac{d\omega_1}{2\pi i} %e^{i\omega_1 0^+} 
    \int \frac{d\omega_2}{2\pi i}
    %e^{i\omega_2 0^+} 
    G_{ik}(\omega + \omega_1) \, G_{kl}(\omega_1 + \omega_2) \, G_{lj}(\omega_2) \, v_{il} \, v_{kj} \\[7pt]
    \nonumber 
    &=&  - \sum_s^{emp}\sum_{m_1}^{occ}\sum_{m_2}^{emp} \frac{A^s_{ik}A^{m_1}_{kl}A^{m_2}_{lj} v_{il}v_{kj}}{\omega - (z_s - z_{m_1} + z_{m_2})} - \sum_s^{occ}\sum_{m_1}^{emp}\sum_{m_2}^{occ}\frac{A^s_{ik}A^{m_1}_{kl}A^{m_2}_{lj} v_{il}v_{kj}}{\omega - (z_s - z_{m_1} + z_{m_2})}.
%\end{align}
\end{eqnarray}
\end{widetext}
In particular from the SOP representation we can directly observe that the direct term of Eq.~\eqref{eq:2bd_sgm} and the exchange term of Eq.~\eqref{eq:2bx_sgm} share exactly the same poles.
However, the residues of the exchange term are all negative semi-definite, while the residues of the direct term are all positive semi-definite. 
The positive semi-definiteness of the 2B self energy depends then on the residues of the direct term being (matrix-wise) larger
than the residues of the exchange term.
Finally the SOP representation of the GW self-energy is given by:
\begin{eqnarray}
    %\label{eq:sgm_GW_c}
    \nonumber
    \Sigma^{GW,c}_{ij}(\omega) &=& - \int \frac{d\omega_1}{2\pi i} e^{i\omega_1 0^+} G_{ij}(\omega + \omega_1)W^c_{ij}(\omega_1)
    \\[7pt]
    \nonumber
    &=& \phantom{-}\sum_s^{occ}\sum_t^{emp} \frac{A^s_{ij}B^t_{ij}}{\omega - (z_s - w_t)} 
    \\ %\nonumber
    \label{eq:sgm_GW_c}
    & & -\sum_s^{emp}\sum_t^{occ} \frac{A^s_{ij}B^t_{ij}}{\omega - (z_s - w_t)}. 
\end{eqnarray}
%
%
% -- SOP SSE --
\section{SSE in the SOP formalism}
\label{sec:sse_sop}
In this Section we report the Sham-Schl\"uter equation written in the SOP formalism. The expression obtained leads to a convenient numerical implementation which does not depend on the explicit form of the self-energy approximation being used.
By expressing all dynamical operators in the coordinate basis on the lattice, the SSE can be rewritten as 
\begin{multline}
    \label{eq:sse_sop}
    \sum_j \int \frac{d\omega}{2\pi i} e^{i\omega 0^+} G^\text{KS}_{ij}(\omega) \, v^{\text{KS}, xc}_{j} \,G_{ji}(\omega) = \\
    %\nonumber 
    =\sum_{jk}\int \frac{d\omega}{2\pi i}e^{i\omega 0^+} G^{KS}_{ij}(\omega)\Sigma_{jk}(\omega)G_{ki}(\omega),  
\end{multline}
where the linearized version can be obtained by just substituting $G\rightarrow G^\text{KS}$ everywhere, including the implicit dependency of the GF of the self-energy.
By introducing the following notation for the left-hand-side (LHS) of Eq.~\eqref{eq:sse_sop}
\begin{equation}
    \label{eq:sse_A}
    A_{ij} = \int \frac{d\omega}{2\pi i} e^{i\omega0^+} \, G^\text{KS}_{ij}(\omega) \, G_{ji}(\omega) 
\end{equation}
%\AFnote{volendo l'exp di convergenza qui si potrebbe togliere (sotto no per colpa si $\Sigma^{HF}$). Uguale in B1.}
and rewriting the right-hand-side (RHS) as 
\begin{equation}
    \label{eq:sse_b}
    b_i = \sum_{jk}\int \frac{d\omega}{2\pi i} e^{i\omega 0^+} \, G^\text{KS}_{ij}(\omega) \,\Sigma_{jk}(\omega) \,G_{ki}(\omega),
\end{equation}
the resulting linear system of equations becomes
\begin{equation}
    \sum_j A_{ij}v^{\text{KS}, xc}_{j} = b_i .
    \label{eq:sse_linsys}
\end{equation}
Since the integrals in frequency of Eqs.~\eqref{eq:sse_A} and~\eqref{eq:sse_b} can be evaluated analytically because of the SOP representations adopted, the expression of the matrix $A_{ij}$ and of the vector $b_i$ can be written explicitly in terms of the poles and residues of the dynamical operators involved.
By adopting the notation:
\begin{align}
    \label{eq:sse_notation}
    G^\text{KS}_{ij}(\omega) &= \sum_s \frac{P^s_{ij}}{\omega - z^\text{KS}_s},\\
    G_{ij}(\omega) &= \sum_s \frac{R^s_{ij}}{\omega - z_s},\\
    \Sigma_{ij}(\omega) = &\Sigma^0_{ij} + \sum_s \frac{\Gamma^s_{ij}}{\omega - \Omega_s}.
\end{align}
we obtain the following expression for the matrix $A_{ij}$:
\begin{equation}
    \label{eq:sse_A_explicit}
    %A^{SSE}_{ij} 
    A_{ij}
    = \sum_s^{emp}\sum_q^{occ} \frac{P^s_{ij}R^q_{ji}}{z_q - z^\text{KS}_s} - \sum_s^{occ}\sum_q^{emp} \frac{P^s_{ij}R^q_{ji}}{z_q - z^\text{KS}_s}. 
\end{equation}
Following Eq.~\eqref{eq:sse_A_explicit}, a possible divergence of the matrix elements of $A$, due to $z_q = z^\text{KS}_s$ can be avoided if $G$ and $G^\text{KS}$ have the same quasiparticle HOMO, since this would avoid that ``empty" poles of one operator overlap with ``occupied" poles of the other. 
Moreover, this alignment of the chemical potentials also makes the linear system defined in Eq.~\eqref{eq:sse_linsys} non rank-defective, since it eliminates the indeterminacy of the potential in terms of a spatially-constant shift.

The entries of the vector $b$ are instead given by:
\begin{eqnarray}
    %\nonumber
    \label{eq:sse_b_explicit}
    %b^{SSE}_i 
    b_i 
    &=& \sum_s^{emp}\sum_q^{occ} \frac{P^s_{ij}\Sigma_{jk}(z_q)R^q_{ki}}{z_q - z^\text{KS}_s}
    \\
    \nonumber
    &-& \sum_s^{occ}\sum_q^{emp} \frac{P^s_{ij}\Sigma_{jk}(z^\text{KS}_s)R^q_{ki}}{z_q - z^\text{KS}_s}\\
    &+& \sum_m^{occ} G^\text{KS}_{ij}(\Omega_m)\,\Gamma^m_{jk}\, G_{ki}(\Omega_m)
    + \sum_s^{occ}\sum_q^{occ} \tilde{b}^{sq}_i . 
    \nonumber
    %\label{eq:sse_b_explicit}
\end{eqnarray}
We note that the expression of the term $\tilde{b}_i^{sq}$ changes according to the presence of a possible pole of second order in the vector $b_i$ of Eq.~\eqref{eq:sse_b}. These are obtained in the case of $z_q = z^\text{KS}_s$, when both poles are below or at the chemical potential.
In this situation the term can be evaluated as 
\begin{equation}
    \label{eq:sse_b_tilde}
    \tilde{b}_i^{sq} = \left\{
    \begin{array}{ll}
        \frac{P^s_{ij}\, \left[\Sigma_{jk}(z_q) - \Sigma_{jk}(z^\text{KS}_s)\right] \,R^q_{ki}}{z_q - z^\text{KS}_s} &\quad \text{if} \quad z_q \neq z^\text{KS}_s \\[7pt]
        P^s_{ij}\, \dot{\Sigma}_{jk}(z_q) \,R^q_{ki}     &\quad \text{if} \quad z_q = z^\text{KS}_s
    \end{array}
    \right.
\end{equation}
where $\dot{\Sigma}(\omega)$ is the frequency derivative of the self-energy.
It is also worth noting that for a self-consistent GF, one has $\Gamma^m G(\Omega_m)=0,\,\,\forall\, m$ since the poles of the self-energy are zeros of the GF (in the subspace spanned by $\Gamma^m$), implying that the first term on the third line of Eq.~\eqref{eq:sse_b_explicit} can be taken to be zero.
%
%
% ======== BIBLIOGRAPHY =========================
% \renewcommand{\emph}{\textit} is needed also for .bbl
% file to have correct formatting 
\renewcommand{\emph}{\textit}
%\bibliographystyle{apsrev4-2}
%\bibliography{biblio.bib}
%======= from .bbl file =========================
% There should be a output.bbl file among the log files of
% overleaf. Copy paste here its content
%======= END PAPER ==============================
% Now start with supplementary
% -- FIGURE COMMANDS --
\renewcommand\thefigure{S\arabic{figure}}
\renewcommand\thesection{S\arabic{section}}
\renewcommand\theequation{\thesection.\arabic{equation}}
\setcounter{section}{0}
\setcounter{figure}{0}
\def\appendixname{}
%
%
%========= SUPPLEMENTARY ========================
%
\clearpage
\title{Supplementary Information: \\
Broken symmetry solutions in one-dimensional lattice models\\ via many-body perturbation theory}
\maketitle
%==========================================
%
% ====== EXACT DIAGONALIZATION
%
\section{Exact diagonalization}
\label{subsec:exact_diago}
The exact ground state and the GF of lattice models can be obtained via exact diagonalization of the many-body Hamiltonian $H_{N_e}$, where $N_e$ is the number of electrons in the system. 
This procedure requires evaluating the matrix elements of the Hamiltonian in the many-body basis and then diagonalizing it, in order to obtain the total energies (both ground and excited states) of the neutral system as eigenvalues and the corresponding many-body wavefunctions as eigenvectors. 
The exact diagonalization becomes rapidly unfeasible with the number of sites in the lattice model.
For a model having $N$ sites, one orbital per site and $N_e$ electrons the number of states in the many-body basis grows as the binomial coefficient  $C_{N_e}^{2N}$:
\begin{equation}
    C_{N_e}^{2N} = \frac{2N !}{\left(2N - N_e\right)! \,N_e!}.
\end{equation}

Importantly, since the exact diagonalization gives access to the ground state many-body wavefunction $| \Psi^0_{N_e}\rangle$ via the eigenvectors of the Hamiltonian, it allows one to evaluate also the expectation value of any combination of field operators on the ground state.
The exact GF can then be built from the result of the exact diagonalization via the evaluation of the eigenenergies of the Hamiltonian projected on the $(N_e - 1)$ and $(N_e + 1)$ sectors of the Fock space:
\begin{eqnarray}
  \nonumber
    G_{ij}(z) &=& \langle \Psi^0_{N_e} | c_{i\sigma} \left[ zI - (H_{N_e + 1} - E^0_{N_e})\right]^{-1} c^\dagger_{j\sigma} | \Psi^0_{N_e}\rangle 
    \\
    &+& \langle \Psi^0_{N_e} | c^\dagger_{j\sigma} \left[ zI - ( E^0_{N_e} - H_{N_e-1})\right]^{-1} c_{i\sigma} | \Psi^0_{N_e}\rangle .
  \nonumber
\end{eqnarray}
In this work we only consider interactions that are diagonal with respect to spin.
overall, the evaluation of the exact GF allows us to determine the exact value of the energy gap and the exact elements of the density matrix, determining its symmetry.  
%
%
% ======== HUBBARD DIMER (AFM) 2B ======
%
\section{AFM solutions in the Hubbard dimer using the 2B self-energy}
\label{app:ssb_hubbard_dimer_2B}
In this Section we report the detailed results about the broken symmetry solutions found in the Hubbard dimer by adopting the 2B self-energy.
In the following we will refer to $G^{HF}$ when considering the fully self-consistent HF solution, while with $G^{trial}$ we consider a trial GF defined by adding the following local and spin-dependent potential $U_{i\sigma}^{trial}$ to the independent particle Hamiltonian:
\begin{equation}
    \label{eq:U_staggered}
    U_{i\sigma}^{trial} = \left\{
    \begin{array}{ll}
        (-1)^i \, U      & \qquad \text{if} \quad  \sigma = \uparrow \\
        (-1)^{i+1} \, U  & \qquad \text{if} \quad \sigma = \downarrow
    \end{array}
    \right.
\end{equation}
where $U$ has the same value of the on-site electron-electron repulsion entering the Hubbard Hamiltonian.

For what concerns the solutions found when using the 2B self-energy, non-magnetic (NM) solutions are obtained by starting the self-consistency loop from a NM $G^{HF}$ and 
we refer to the total energy evaluated in this way as $E_{NM}$, while AFM solutions can be found by starting the loop from an unrestricted symmetry (US) GF, which can be either an antiferromagnetic (AFM) $G^{HF}$ or an AFM $G^{trial}$, and in this case we refer to the total energy as $E$.
Notably, we observe that 
a loop starting from a US GF can converge either to NM or to AFM solutions. 
In the latter case the symmetry of the final solution can be determined by evaluating the order parameter $M^{LRO}$ of the main text.

\begin{figure}
    \centering
    \includegraphics[width=0.45\textwidth]{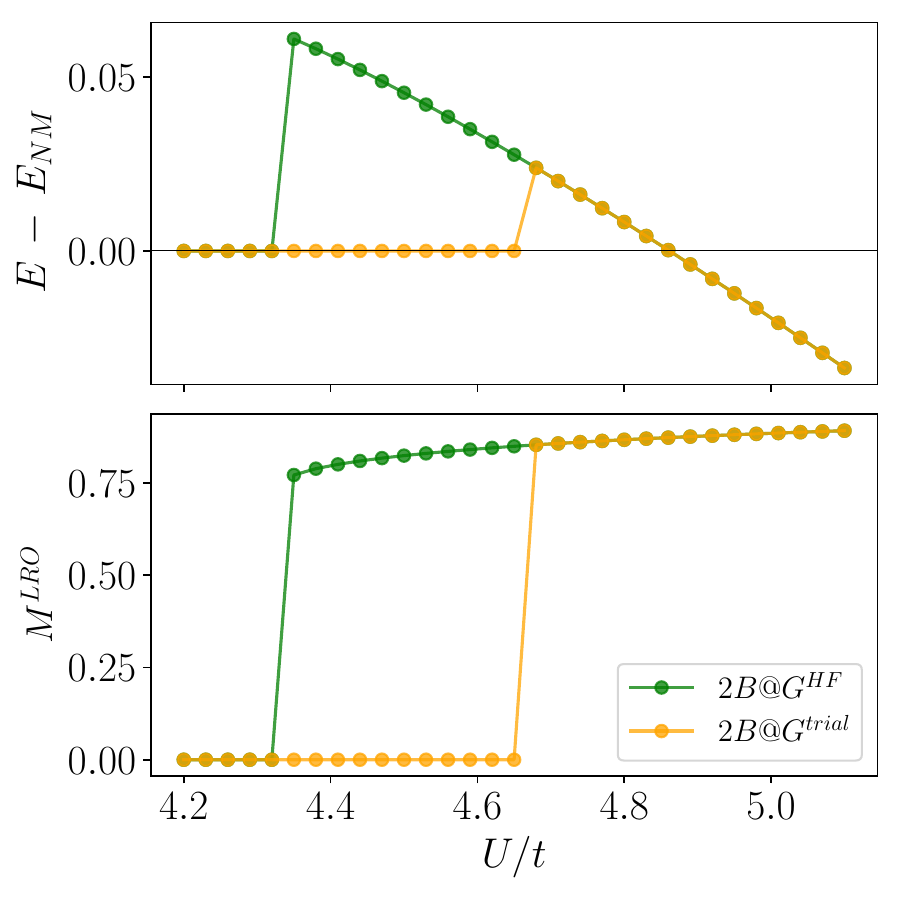}
    \caption{Study of the AFM    symmetry breaking in the Hubbard dimer treated using the 2B self-energy. For each value of the parameter $U/t$ we initialize the self-consistency loop with a non-magnetic (NM) GF and with an unrestricted symmetry (US) GF.
    In the upper panel we report the difference $E - E_{NM}$ between the two corresponding energies. 
    In the lower panel we report the order parameter evaluated on the unrestricted symmetry solution of the self-consistent loop.
    We employ US GFs obtained either from a HF calculation or via the trial potential given in Eq.~\eqref{eq:U_staggered}.
    \label{fig:ssb_comp_hubbard_dimer_2B} \\
    }
\end{figure}

In Fig.~\ref{fig:ssb_comp_hubbard_dimer_2B} we report the energy difference between $E$ and $E_{NM}$ and the order parameter $M^{LRO}$ of the respective $GF$s.
We observe that, in the case of the 2B self-energy, there is an interval of values ($U/t < 4.83$) where we can converge to either an AFM or a NM solution depending on the starting point. 
In this interval, however, the AFM solution is not the ground state, while it becomes the ground state only for interaction strengths larger than $U/t \simeq 4.83$.
Whenever we converge to an AFM solution when starting from an US $G^{trial}$ we also obtain the same order parameter and the same energy of the AFM solution obtained by starting from an US $G^{HF}$. 
This signals that we find the same AFM solution in both cases and multiple AFM branches are not found within our calculations.
We also notice that the order parameter of the US solutions does not go continuously to zero with $U/t$.

\begin{figure*}[]
    \centering
    \includegraphics[width=0.8\textwidth]{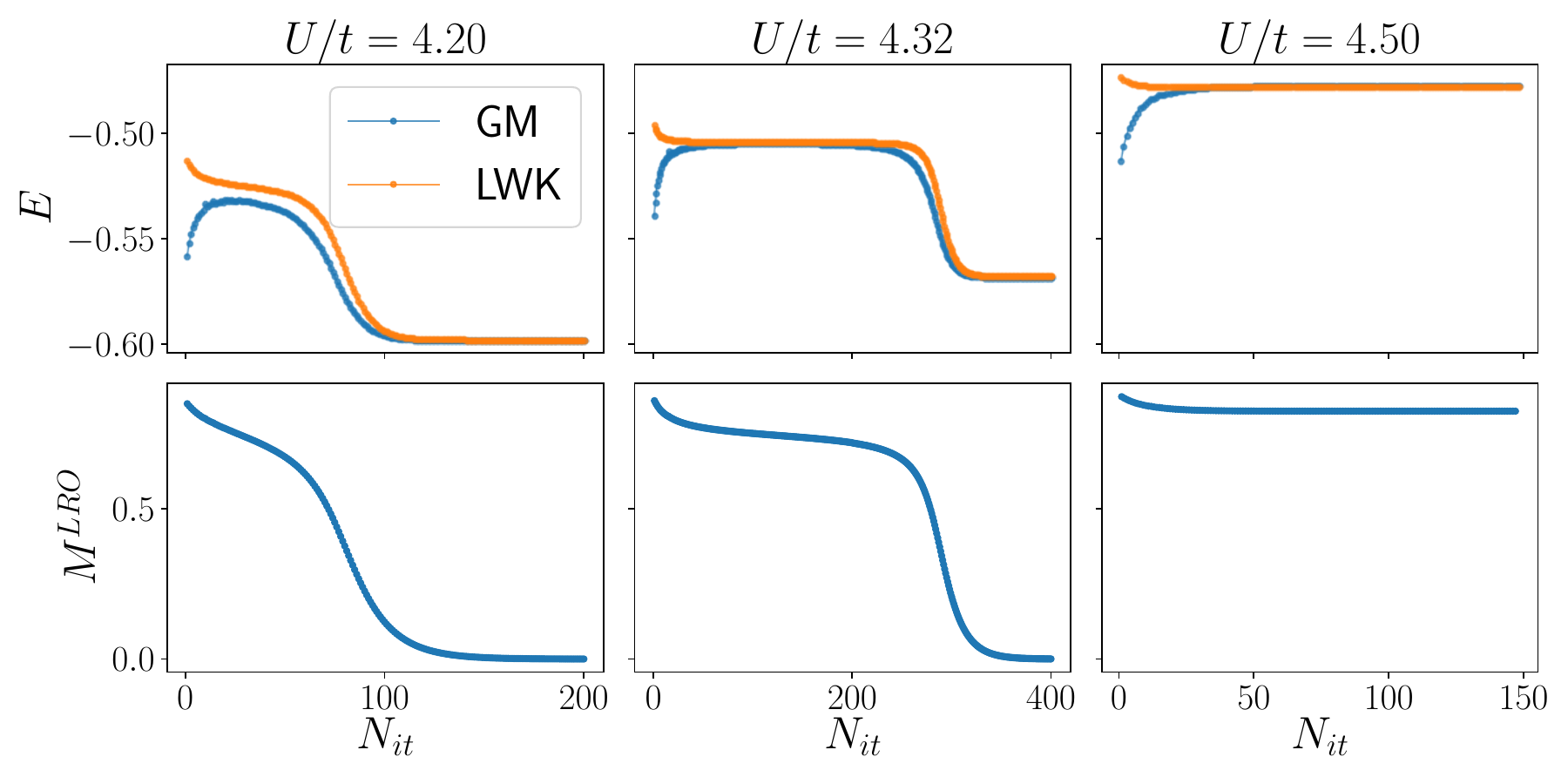}
    \caption{Run values of the total energy (first row) and order parameter $M^{LRO}$ (second row) for the half-filled Hubbard dimer studied with the 2B. The total energy is calculated both with the Galitskii-Migdal (GM) formula and the Klein (LWK) functional. 
    The number of iterations is reported with a logarithmic scale.
    \label{fig:ssb_run_2B}}
\end{figure*}

In Fig.~\ref{fig:ssb_run_2B} we show how the total energy and the order parameter change during the self-consistency loop starting with the US $G^{HF}$.
The total energy has been evaluated both with the Galitskii-Migdal (GM) formula and with the Klein functional (LWK).
The values of the GM and LWK total energies agree at self-consistency, as expected.
We notice that at smaller values of the interaction strength ($U/t = 4.20$), the first steps within the loop are characterized by the presence of states that are near in energy, but having significantly different values of the order parameter. 
The loop then eventually converges towards a NM state at lower energy.

Around $U/t\simeq4.32$, which is the value for which the AFM solutions are found when starting from the US $G^{HF}$, the Klein functional shows a flattening of the states found at the beginning of the loop. 
However, these states are still characterized by different values of the order parameter and the stopping condition of the loop, based on the density matrix results in the convergence to a NM state at lower energy.
At the larger interaction strength of $U/t = 4.50$, the self-consistency loop converges towards the AFM state found at the beginning, which is now characterized by a well defined total energy and order parameter.
It is important to notice, however, that for $U/t = 4.50$ the AFM state found is not the ground state.

We also notice that in any case the order parameter gets reduced during the self-consistency loop with respect to its initial value.
We can conclude that the convergence towards an AFM solution starting from a US $G^{HF}$ or from $G^{trial}$ depends on the order parameter at the beginning of the loop being sufficiently large.
We also notice that the presence of plateaux in the total energy during the self-consistency loop has been observed also in Ref.~\cite{Joost2021ContrPlasmaPhys--suppl}, where the 2B self-energy is also employed in the study of broken symmetry solutions. 
We note that within all the self-energy approximations considered in this work, 2B is actually the one giving the larger value of $U/t$ for which the ground state is predicted to have an AFM symmetry and so it is the one predicting the correct NM symmetry for the largest interval of values of $U/t$.
A possible explanation can be related to the fact that the second order exchange diagram contained within the 2B self-energy is able to include the effect of a partial vertex correction~\cite{Romaniello_hubGW_2009J.Chem.Phys--suppl}, which is relevant in the description of the physics of Hubbard systems.
%
%
%========= HUBBARD CHAIN NM =================
\section{Nonmagnetic Hubbard chains}
\label{sec:Hubbard_chains_PM}
\begin{figure}
    \centering
    \includegraphics[width=0.45\textwidth]{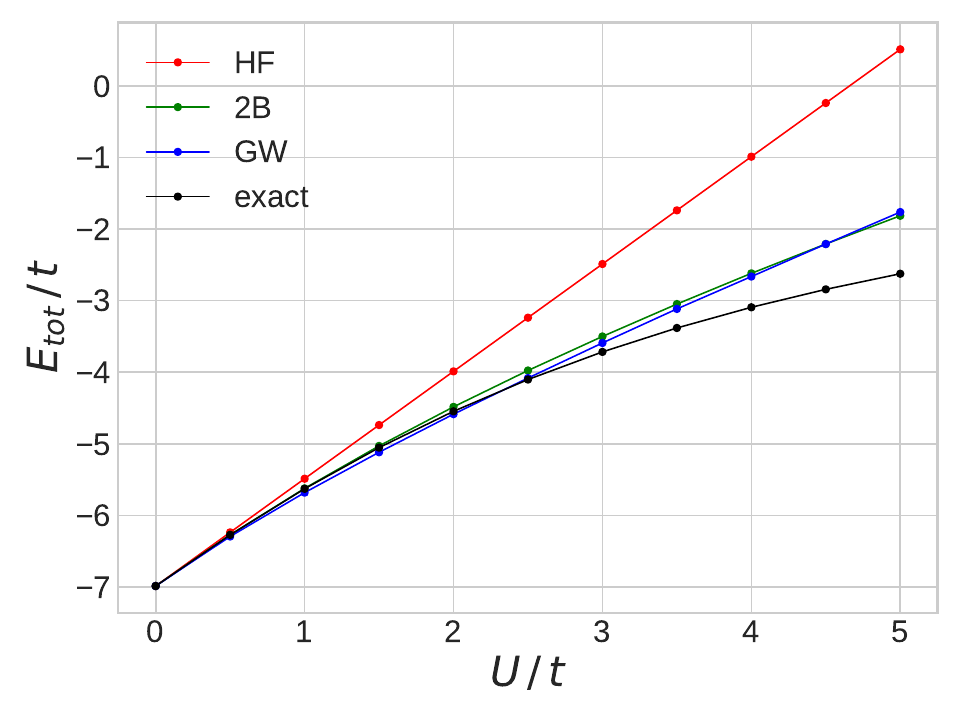}
    \caption{Total energy of a $N=6$ Hubbard chain with open boundary condition as a function of the ratio $U/t$ for different self-energy approximations. The value of the total energy is scaled with the hopping parameter $t$.
    The value of $E_{tot}$ reported here is the one evaluated through the Klein functional.
    \label{fig:hub_N6_energy}}
\end{figure}

In this Section we provide detailed results about the nonmagnetic (NM) solutions in Hubbard chains.
In Fig.~\ref{fig:hub_N6_energy} we report the total energy for a $N=6$ Hubbard chain with open boundary conditions and for different values of $U/t$ in the range $U/t\in\left[0,5\right]$ and for all the different self-energy approximations considered. 
The exact results are also shown.
For the different self-energy approximations, the energy values reported are those evaluated via the Klein functional and according to Ref.~\cite{Ferretti2024PRB--suppl}.
We observe that the total energy in the case of HF increases linearly with $U/t$, as expected in the case of NM solutions in Hubbard clusters~\cite{Stan2015NewJPhys--suppl}. 
The 2B and GW self-energies give instead similar values of the total energy in the whole interval of $U/t$ considered, while significant deviations from the exact result are found for $U/t > 3$.

\begin{figure}
    \centering
    \includegraphics[width=0.45\textwidth]{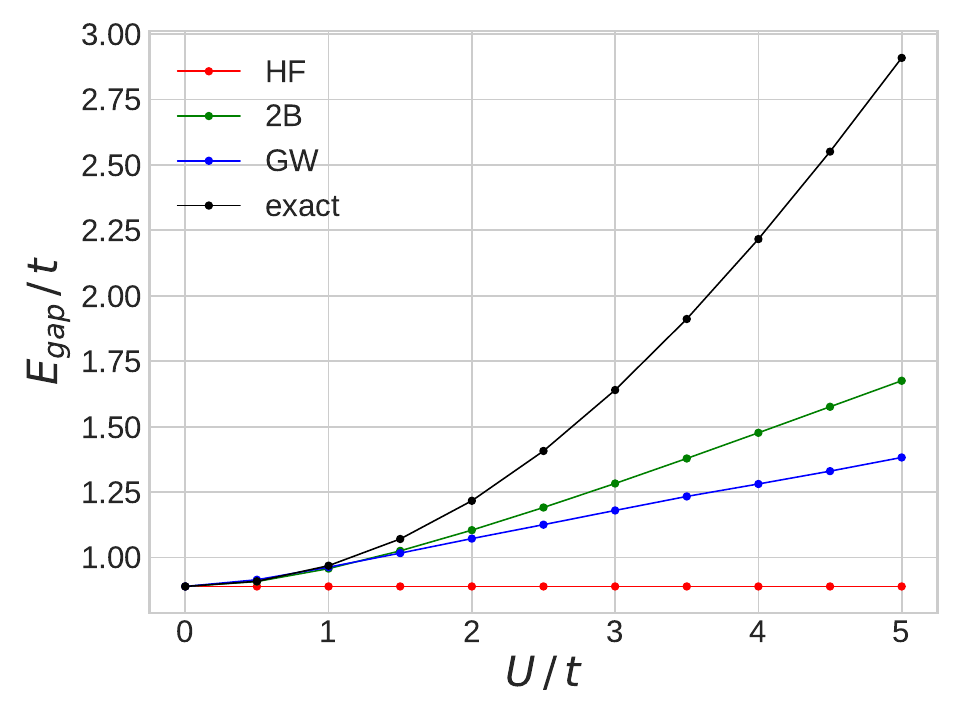}
    \caption{Energy gap of a $N=6$ Hubbard chain with open boundary condition as a function of the ratio $U/t$ for different self-energy approximations. The value of the energy gap is scaled with the hopping parameter $t$.
    \label{fig:hub_N6_gap}}
\end{figure}

In Fig.~\ref{fig:hub_N6_gap} we report the energy gap for the $N=6$ Hubbard chain described previously.
Here we notice that in the case of HF, the gap does not increase with respect to the non-interacting case when $U/t$ is increased.
This is due to the fact that for the Hubbard model in the NM ground state and at half-filling the HF self-energy reduces trivially to a constant and $U$-dependent energy correction, leading to an overall rigid shift of the total energy given by $\Delta E^{Hub,NM}_{HF} = UN/4$, with $N$ the number of sites in the chain.   

The 2B and GW self-energies show instead an increase in the energy gap with $U/t$, which is due to the inclusion of correlation effects beyond HF. 
In particular we observe that 2B gives a larger gap with respect to the GW in the case of strong correlation.
However, the energy gap found via approximate MBPT self-energies in the strong correlation regime is severely underestimated with respect to the exact one.

\begin{figure}
    \centering
    \includegraphics[width=0.45\textwidth]{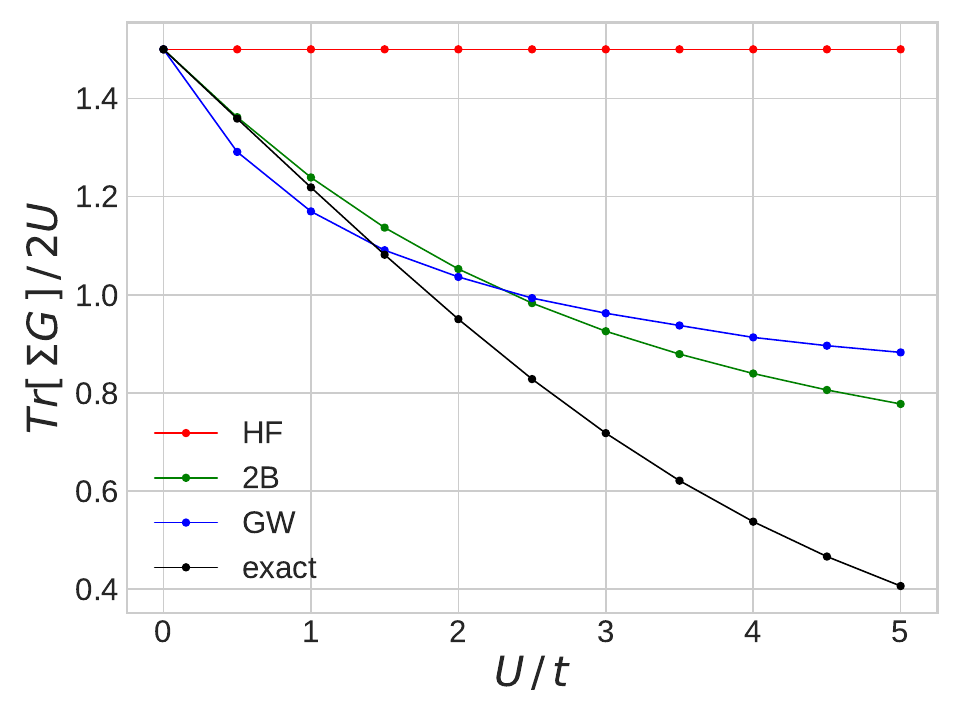}
    \caption{Values of $D$ from Eq.~\eqref{eq:double_occupancy} of a $N=6$ Hubbard chain with open boundary condition as a function of the ratio $U/t$ for different self-energy approximations.}
    \label{fig:hub_N6_double}
\end{figure}

Another quantity of interest in the Hubbard model is represented by the double occupancy $\langle n_{i\uparrow} n_{i\downarrow}\rangle$ correlation function, which gives an estimate of the antiferromagnetic (AFM) correlations within the system -- the lower the expectation value, the higher the AFM correlations included in the ground state.
In particular the quantity $D := \sum_i \langle n_{i\uparrow } n_{i\downarrow}\rangle$ can be easily related to the ground state total and kinetic energies of the Hubbard Hamiltonian:
\begin{equation}
    \label{eq:double_occupancy}
    D := \sum_i \langle n_{\uparrow i} n_{\downarrow i}\rangle = \frac{E_{tot} - \Trw\left\{h_0 G\right\}}{U} = \frac{\Trw\left\{\Sigma G\right\}}{2U},
\end{equation}
%with the kinetic energy given by $E_{kin} = \Trw\left\{h_0 G(\omega)\right\}$ and
where the last equality of Eq.~\eqref{eq:double_occupancy} comes from the Galitskii-Migdal formula~\cite{GalitskiiMigdal_SovPhys_1958--suppl}. 
While in our work Eq.~\eqref{eq:double_occupancy} is used to compare MBPT results to the exact ones, a detailed discussion about its validity to describe the double occupancy can be found in Ref.~\cite{farid2021luttingerwardfunctionalconvergenceskeleton--suppl}. 
In Fig.~\ref{fig:hub_N6_double} we report the values of $D$ for the $N=6$ Hubbard chain while varying $U/t$.
We observe that antiferromagnetic correlations are only included in the 2B and GW levels of theory for the NM ground state. The value of $D$ is however larger than what found via an exact treatment. 
%
%
% ====== HUBBARD CHAIN (AFM) ==========
%
\section{Broken symmetry solutions in longer Hubbard chains}
\label{sec:ssb_hubbard_long}
\begin{figure}
    \centering
    \includegraphics[width=0.45\textwidth]
    {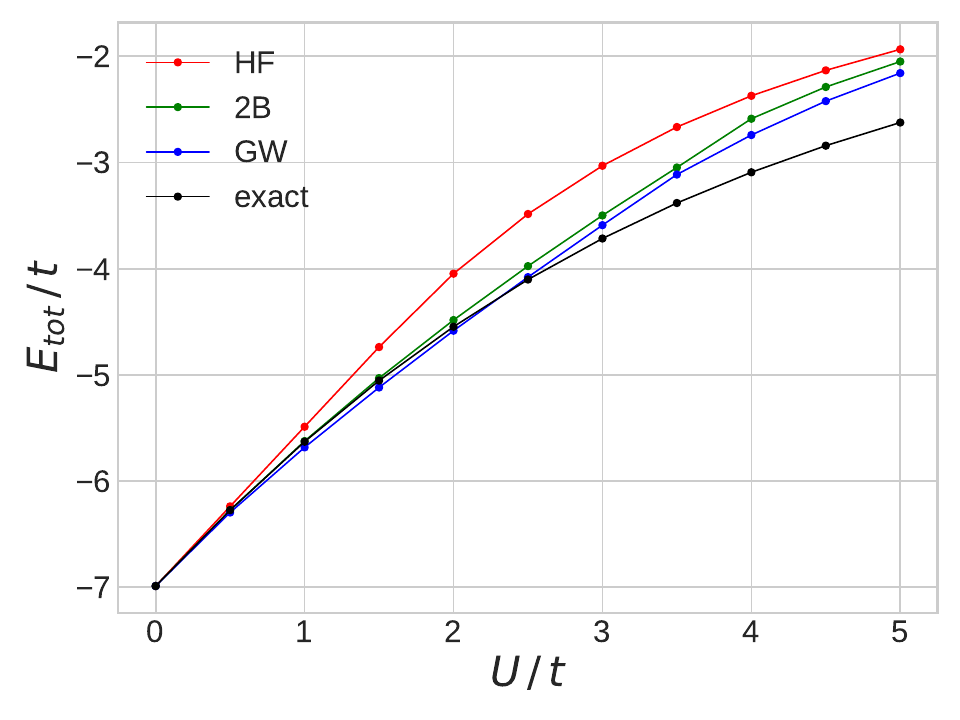}
    \caption{Total energy of a $N=6$ Hubbard chain at half-filling and with open boundary conditions for self-consistent GF obtained starting the loop from a trial US GF. Different self-energy approximations are considered and the results are compared with the exact ones, obtained from an exact diagonalization treatment.
    }
    \label{fig:ssb_hubbard_N6_energy}
\end{figure}

In this Section we discuss in detail the antiferromgnetic (AFM) broken symmetry solutions found in longer Hubbard chains.
We focus on the case of the $N=6$ half-filled Hubbard chain with open boundary conditions.
We look for broken symmetry solutions by allowing the self-consistency loop to start from an unrestricted symmetry (US) trial GF. 
It is important to stress that starting from a US GF is a necessary condition to find a broken symmetry self-consistent GF, but not a sufficient one. 
In Fig.~\ref{fig:ssb_hubbard_N6_energy} we report the total energy of the $N=6$ chain. 
Besides different self-energy approximations, we also report the exact results obtained through exact diagonalization.
We observe that total energy curves evolve in a smooth way with respect to the parameter $U/t$ for all the self-energy approximations considered.
In the range of parameters considered, 2B and GW give similar values of the total energy.

\begin{figure}
    \centering
    \includegraphics[width=0.45\textwidth]{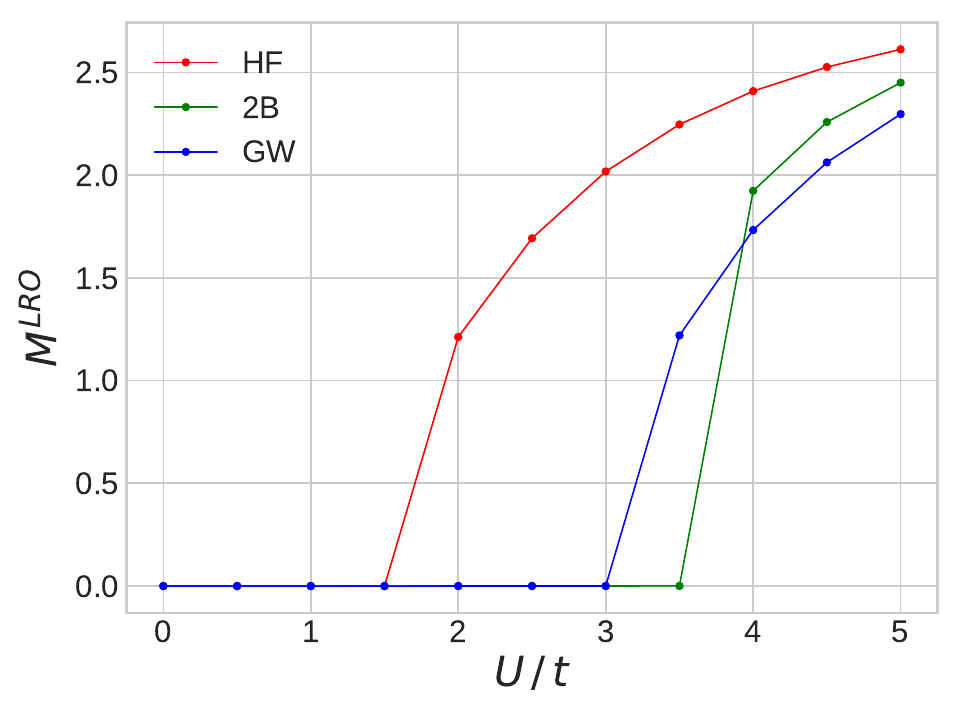}
    \caption{Order parameter $M^{LRO}$, for a $N=6$ Hubbard chain at half-filling and with open boundary conditions for the self-consistent GF obtained starting the loop from a trial US GF. 
    The exact result is not reported, as the exact ground state shows an identically vanishing order parameter.}
    \label{fig:ssb_hubbard_N6_lro}
\end{figure}

\begin{figure}
    \centering
    \includegraphics[width=0.45\textwidth]{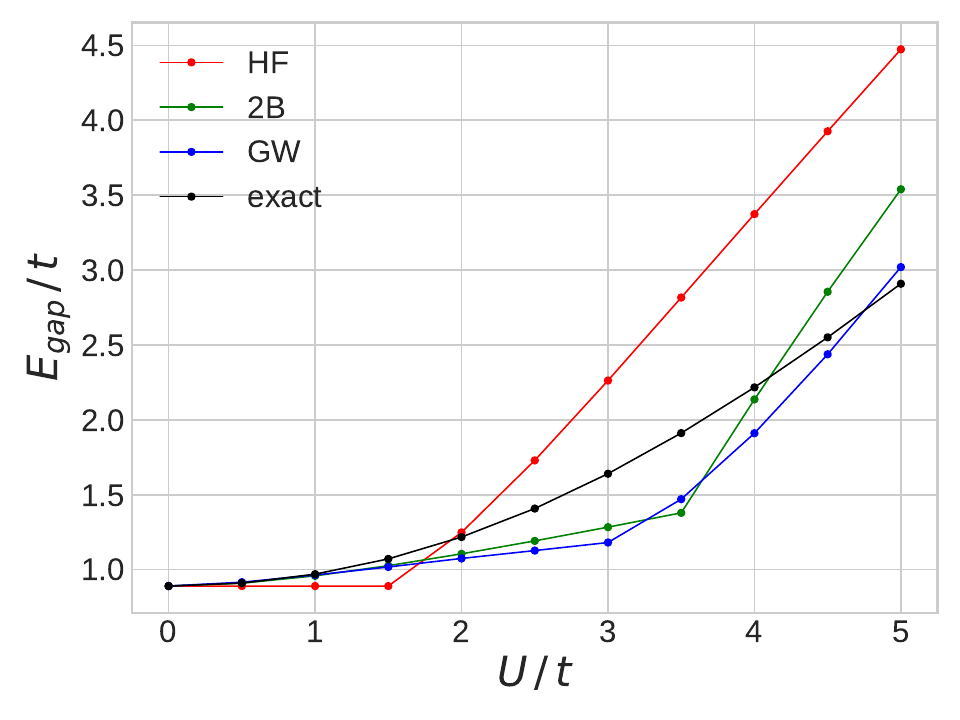}
    \caption{Energy gap of a $N=6$ Hubbard chain at half-filling and with open boundary conditions for self-consistent GF obtained starting the loop from a trial US GF.
    Different self-energy approximations are considered and the results are compared with the exact ones, obtained from an exact diagonalization treatment.}
    \label{fig:ssb_hubbard_N6_gap}
\end{figure}

Since it is important to understand whether the self-consistent solution is AFM or not, in Fig.~\ref{fig:ssb_hubbard_N6_lro} we report the corresponding values of the order parameter $M^{LRO}$, evaluated according to the main text.
We observe that the different self-energy approximations start converging towards AFM solutions at different values of the parameter $U/t$. 
The 2B approximation is the one that predicts the correct NM symmetry of the ground state for the largest interval of values of $U/t$.
All the self-energy approximations also approach similar values of $M^{LRO}$ in the region of larger values of $U/t$.
In Fig.~\ref{fig:ssb_hubbard_N6_gap} we report the value of the energy gap for the $N=6$ Hubbard chain.
We observe that, for each self-energy approximation, as soon as the AFM solution is reached the energy gap shows a significant increase, becoming then more similar to the exact result.
However, in the large $U/t$ regime, 2B and GW ultimately reach values of the energy gap that are larger than the exact ones.

\begin{figure}
    \centering
    \includegraphics[width=0.45\textwidth]{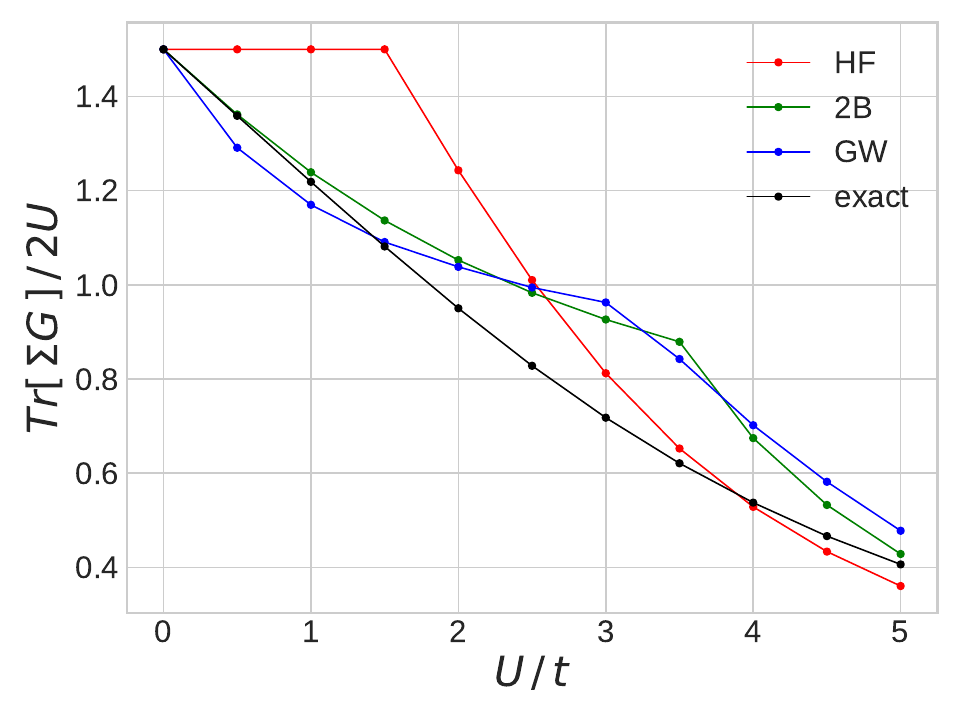}
    \caption{Double occupancy evaluated from Eq.~\eqref{eq:double_occupancy} for a $N=6$ Hubbard chain at half-filling and with open boundary conditions for self-consistent GF obtained starting the loop from a trial US GF.
    Different self-energy approximations are considered and the results are compared with the exact ones, obtained from an exact diagonalization treatment.
    Energies are scaled with the parameter $t$.}
    \label{fig:ssb_hubbard_N6_double}
\end{figure}

In Fig.~\ref{fig:ssb_hubbard_N6_double} we report the value of the double occupancy $D$, evaluated from Eq.~\eqref{eq:double_occupancy}, for the $N=6$ Hubbard chain.
The exact solution of the Hubbard model features the presence of large AFM correlations in the NM ground state. 
Reaching larger values of the correlation strength $U/t$, the approximate self-energies can not introduce such a large AFM correlation in their NM ground state. 
The AFM unphysical symmetry breaking could then be seen as an artifact through which even an approximate self-energy manages to include larger AFM correlations in the large $U/t$ regime. 
This can be related to the fact that all the AFM solutions considered approach the exact result in the large correlation regime.
It is also interesting to notice that the best agreement is obtained for the 2B level of theory, both in the small and in the large correlation regime.

\begin{figure}[]
    \centering
    \includegraphics[width=0.45\textwidth]{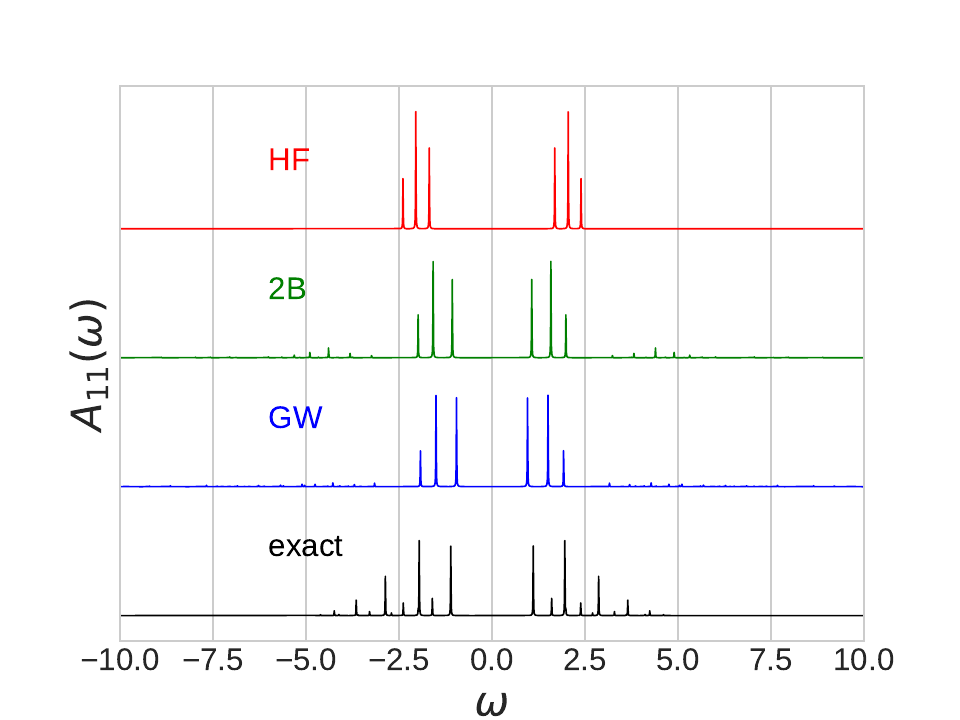}
    \caption{Spectral density $A_{11}(\omega)$ on the first site of an open Hubbard chain with $N=6$ and $U/t=4$. The value of $\omega$ is scaled with the hopping term $t$ and the chemical potential has been shifted to $\omega=0$. The spectral function is reported in arbitrary units. The exact solution is also reported. The plot has been obtained by including a finite imaginary shift $\eta=5\times 10^{-3}$ in the poles of the GF.}
    \label{fig:hub_N6_U4_spectral_ssb}
\end{figure}

Next, in Fig.~\ref{fig:hub_N6_U4_spectral_ssb} we consider how the spectral function changes in the case of broken symmetry solutions.
\vspace{70pt}
In this case we achieve a better agreement between the spectral functions evaluated via approximate MBPT and the exact result. 
%
%
%====== METALLIC NON-INTERACTING STATE =====
%
\section{Independent particle energies of a non-interacting closed chain}
\label{app:metal_N4}
In the case of a non-interacting system the lattice Hamiltonian simply reduces to the following tight-binding model:
\begin{equation}
    H_0 = -t \sum_{\sigma}\sum_{\langle i j \rangle} \left( c^\dagger_{i\sigma} c_{j\sigma} + h.c. \right).
\end{equation}
In this situation the independent-particle picture is exact and the total energy can be computed as the sum of independent-particle energies.
The independent-particle eigenvalues in a ring of $N$ sites are given by:
\begin{equation}
    \epsilon_n = 2t \text{cos}\left(\frac{2\pi n}{N}\right),
\end{equation}
with the integer index $n$ ranging from $0$ to $(N-1)$.
The energy range spanned by the $\epsilon_n$ is fixed between $[-2t,2t]$ and at zero temperature it is always possible to choose the Fermi energy as $E_F = 0$.
In particular if $N$ is an integer multiple of $4$, so that $N=4M$, with $M$ an integer such that $M\ge 1$, then there are two indexes $m_1,m_2$ such that:
\begin{align}
    m_1 &= M \\
    m_2 &= 3M \\
    \epsilon_{m_1} &= \epsilon_{m_2} = 0.
\end{align}
In particular as only $2M$ states can be occupied in a half-filled chain, then only one of the states out of $\epsilon_{m_1}$ and $\epsilon_{m_2}$ can be occupied, describing a metallic system.
%
%
%============================
% BIBLIOGRAPHY
\renewcommand{\emph}{\textit}
%\bibliographystyle{apsrev4-2}
%============================
% Insert here output from .bbl file
\end{document}